\documentclass[11pt,A4paper]{article}
\pdfoutput=1
\usepackage{jcappub}
\usepackage{subcaption}

\begin{document}

\newcommand{\kin}{\alpha_{\rm K}}
\newcommand{\run}{\alpha_{\rm M}}
\newcommand{\bra}{\alpha_{\rm B}}
\newcommand{\ten}{\alpha_{\rm T}}
\newcommand{\beh}{\alpha_{\rm H}}
\newcommand{\hkin}{\hat{\alpha}_{\rm K}}
\newcommand{\hrun}{\hat{\alpha}_{\rm M}}
\newcommand{\hbra}{\hat{\alpha}_{\rm B}}
\newcommand{\hten}{\hat{\alpha}_{\rm T}}
\newcommand{\hbeh}{\hat{\alpha}_{\rm H}}
\newcommand{\trhom}{\tilde{\rho}_{\rm m}}
\newcommand{\tpm}{\tilde{p}_{\rm m}}
\newcommand{\trhos}{\tilde{\rho}_{\rm s}}
\newcommand{\tps}{\tilde{p}_{\rm s}}
\newcommand{\rhom}{\rho_{\rm m}}
\newcommand{\pmm}{p_{\rm m}}
\newcommand{\rhos}{\rho_{\rm s}}
\newcommand{\ps}{p_{\rm s}}
\newcommand{\rhomc}{\rho_{\rm mc}}
\newcommand{\pmc}{p_{\rm mc}}
\newcommand{\rhosc}{\rho_{\rm sc}}
\newcommand{\psc}{p_{\rm sc}}
\newcommand{\Ms}{M_*^2}
\newcommand{\cs}{c_{\rm s}^2}
\newcommand{\csn}{c_{\rm sN}^2}

\newcommand{\DTrem}[1]{\textcolor{blue}{\textbf{[DT: #1]}}}
\newcommand{\DT}[1]{\textcolor{blue}{#1}}
\newcommand{\EBrem}[1]{\textcolor{red}{\textbf{[EB: #1]}}}
\newcommand{\EB}[1]{\textcolor{red}{#1}}
\newcommand{\PGFrem}[1]{\textcolor{green}{\textbf{[EB: #1]}}}
\newcommand{\PGF}[1]{\textcolor{green}{#1}}

\title{The phenomenology of beyond Horndeski gravity}
\author[a]{Dina Traykova,}
\author[a]{Emilio Bellini,}
\author[a]{Pedro G.\ Ferreira}
\affiliation[a]{Astrophysics, University of Oxford, DWB, Keble Road, Oxford OX1 3RH, UK}
\emailAdd{dina.traykova@physics.ox.ac.uk}
\emailAdd{emilio.bellini@physics.ox.ac.uk}
\emailAdd{pedro.ferreira@physics.ox.ac.uk}

\abstract{
We study the phenomenology of the beyond Horndeski class of scalar-tensor theories of gravity, which on cosmological scales can be characterised in terms of one extra function of time, $\beh$,  as well as the usual four Horndeski set of free functions.
We show that $\beh$ can be directly related to the damping of the matter power spectrum on both large and small scales. 
We also find that the temperature power spectrum of the cosmic microwave background (CMB) is enhanced  at low multipoles and the lensing potential is decreased, as a function of $\beh$.  
For a particular choice of time dependence we find constraints on $\beh$ of order ${\cal O}(1)$ using measurements of the temperature and polarisation of the CMB, as well as the lensing potential derived from it, combined with large scale structure data.
We find that redshift space distortion measurements can play a significant role in constraining these theories.
Finally, we comment on the implications of the recent observation of an electromagnetic counterpart to a gravitational wave signal;
we find that these measurements reduce the number of free parameters of the model but do not significantly change the constraints on the remaining parameters.
}


\maketitle

\section{Introduction}\label{sec:intro}

In the last two decades, there has been tremendous progress in testing gravity on cosmological scales.
We can now measure the expansion history, the Cosmic Microwave Background (CMB) and the Large Scale Structure (LSS) of the universe with incredible precision. 
Distance measurements from supernovae \cite{Riess:2006fw,Kowalski:2008ez,Perlmutter:1998np}, combined with measurements of LSS clustering \cite{Springel:2006vs,Anderson:2013zyy,2011MNRAS.418.1707B} and the CMB \cite{Spergel:2003cb,Bennett:2003bz,Spergel:2006hy,Dunkley:2008ie,Ade:2013zuv,Ade:2013sjv, Ade:2015xua, Adam:2015rua} have confirmed that, at recent times, the universe is undergoing a phase of accelerated expansion. 
The simplest model that can explain these measurements  is $\Lambda$CDM, in which the energy density content of the Universe is dominated by an additional stress-energy component in the form of a cosmological constant ($\Lambda$) that causes the observed acceleration. 

The standard model of cosmology assumes that the theory of gravity is General Relativity (GR); indeed the current assumption is that GR  works incredibly well over 15 orders of magnitude in length scale \cite{Baker:2014zba}.
However, the large amount of fine-tuning required to explain the value of the cosmological constant has led to the exploration of alternative models, from Dark Energy (DE) models, where an additional dynamical degree of freedom (d.o.f.) replaces $\Lambda$, to theories that directly modify the laws of gravity, i.e.~Modified Gravity (MG) models \cite{Clifton:2011jh,Kunz:2012aw}. 
A particular class of DE/MG models are the  scalar-tensor theories, where the additional d.o.f.~is represented by a scalar field.

The Horndeski class of theories \cite{Horndeski:1974wa, Deffayet:2011gz,Kobayashi:2011nu}  has received a lot of attention as it is given by a general scalar-tensor action that leads to at most second-order equations of motion on any background and also satisfies the weak equivalence principle, i.e.~all matter (except from the scalar, $\phi$) is coupled minimally and universally to the same metric.
In these theories, there is one extra propagating degree of freedom, $\phi$, in addition to the standard two tensor modes of GR.
It includes, as special cases, a wide variety of dark energy and modified gravity models, such as quintessence \cite{PhysRevD.37.3406}, $f(R)$, Brans-Dicke \cite{PhysRev.124.925}, kinetic gravity braiding \cite{Deffayet:2010qz}, covariant galileons \cite{Deffayet:2009wt}, Chameleons \cite{Khoury:2003aq, Brax:2004qh} and others (for a review see \cite{Clifton:2011jh,Kobayashi:2019hrl}).

It has been shown that Horndeski is not as general as previously believed and can be further extended to the so-called beyond Horndeski theories \cite{Gleyzes:2013ooa,Zumalacarregui:2013pma}.
While in general beyond Horndeski have higher (than second) order equations of motion, this does not necessarily lead to Ostrogradski instabilities \cite{Zumalacarregui:2013pma,deRham:2011qq,Gleyzes:2014dya}.
Indeed, it is possible to show that the time-derivatives remain second-order in the equations of motion for the propagating degrees of freedom for any ADM decomposition \cite{Gleyzes:2014qga,Lin:2014jga,Crisostomi:2016tcp}.
As expected, beyond Horndeski theories lead to new interesting phenomenological properties that are not described by the Horndeski class \cite{PhysRevLett.114.211101}.
For example, a  recent  study \cite{Babichev:2018rfj} shows that the scalar field in beyond Horndeski theories can significantly modify the speed of sound in the atmosphere of the Earth.
The question of whether this set of theories is viable for explaining dark energy has been studied extensively in \cite{Crisostomi:2017pjs,Lim:2010yk,Gao:2014soa};
self-acceleration solutions exist that can lead to a de Sitter universe at late times without the need of a cosmological constant.

The aim of this work is to explore the phenomenology of beyond Horndeski models and constrain the free parameters of the theory using the latest CMB and large scale structure data.
Such fits to data have been performed previously on specific modified gravity models \cite{Barreira:2013jma,Dossett:2014oia,Raveri:2014cka} or specific data sets \cite{Hu:2015rva,Ade:2015rim}.
More generally this was done for the Horndeski models considering a collection of latest CMB and large scale structure data in \cite{Bellini:2015xja,Kreisch:2017uet,Noller:2018wyv}.

We use the formalism introduced in \cite{Bellini:2014fua} for Horndeski and used in \cite{Bellini:2015wfa, Bellini:2015xja, Zumalacarregui:2016pph,Alonso:2016suf} and extended to beyond Horndeski in \cite{Gleyzes:2014rba} (see also \cite{Gubitosi:2012hu,Bloomfield:2012ff,Gleyzes:2013ooa, Gleyzes:2014qga,Lagos:2017hdr} for a different approach to the linear perturbation theory in the Horndeski and beyond Horndeski Lagrangian).
In this approach the evolution of linear perturbations is described fully by the fraction of matter density today, the Hubble parameter $H(t)$ and five other functions of time.
We have chosen the parametrisation for these five functions proposed in \cite{Bellini:2014fua}, in which these are proportional to the fractional energy density of dark energy, $\Omega_{\rm DE}(t)$.
In this way the effective freedom of the model is reduced to just five free parameters.

The structure of this work is as follows. In Section~\ref{sec:theory} we review the theoretical properties of the beyond Horndeski action and the equations of motion that it leads to.
Then in Section~\ref{sec:phenomenology} we discuss the phenomenology of this model.
We present our results for the constraints on the parameters of this model with data in Section~\ref{sec:results}. Finally we discuss our findings in Section~\ref{sec:discussion}.

\section{Theoretical framework}\label{sec:theory}
In this section we discuss the general properties of the beyond Horndeski class of theories, present its Lagrangian, background equations and stability.
We then briefly discuss the recent restrictions arising from GW170817 and finally present the linear perturbation equations for the model. The action for beyond Horndeski reads \cite{Gleyzes:2014dya,Zumalacarregui:2013pma},
\begin{equation}
S[g_{\mu\nu},\phi]=\int\mathrm{d}^{4}x\,\sqrt{-g}\Bigl[\sum_{i=2}^{5}\frac{1}{8\pi G_{\text{N}}}{\cal L}_{i}(g_{\mu\nu},\phi) +\mathcal{L}_{\text{m}}(g_{\mu\nu},\psi_{M})\Bigr]\,,\label{equ:action}
\end{equation}
where $g_{\mu\nu}$ is the metric, $g$ is its determinant and  $\mathcal{L}_{\text{m}}$ generically describes all matter species, which we consider to be minimally and universally coupled to the metric $g_{\mu\nu}$. The sum is over four Lagrangians defined as,
\begin{equation}
\begin{aligned}
{\cal L}_{2} = &\, G_{2}(\phi,\, X)\,,\\
{\cal L}_{3} = &\, -G_{3}(\phi,\, X)\Box\phi\,,\\
{\cal L}_{4} = &\, G_{4}(\phi,\, X)R+G_{4X}(\phi,\, X)\left[\left(\Box\phi\right)^{2}-\phi_{;\mu\nu}\phi^{;\mu\nu}\right]\\
& + F_4(\phi,\, X)\epsilon^{\mu\nu\rho}_{\;\;\;\;\;\;\;\sigma}\epsilon^{\mu^{\prime}\nu^{\prime}\rho^{\prime}\sigma}\phi_{;\mu}\phi_{;\mu^{\prime}}\phi_{;\nu\nu^{\prime}}\phi_{;\rho\rho^{\prime}}\,,\\
{\cal L}_{5} = &\, G_{5}(\phi,\, X)G_{\mu\nu}\phi^{;\mu\nu}\\
& - \frac{1}{6}G_{5X}(\phi,\, X)\Bigl[\left(\Box\phi\right)^{3}+2{\phi_{;\mu}}^{\nu}{\phi_{;\nu}}^{\alpha}{\phi_{;\alpha}}^{\mu}-3\phi_{;\mu\nu}\phi^{;\mu\nu}\Box\phi\Bigr]\\
& + F_5(\phi,\, X)\epsilon^{\mu\nu\rho\sigma}\epsilon^{\mu^{\prime}\nu^{\prime}\rho^{\prime}\sigma^{\prime}}\phi_{;\mu}\phi_{;\mu^{\prime}}\phi_{;\nu\nu^{\prime}}\phi_{;\rho\rho^{\prime}}\phi_{;\sigma\sigma^{\prime}}\,,\label{equ:lag}
\end{aligned}
\end{equation}
where $\phi_{;\mu}\equiv\nabla_\mu\phi$, $\phi_{;\mu\nu}\equiv\nabla_\mu\nabla_\nu\phi$, $G_i(\phi,\, X)$ and $F_i(\phi,\, X)$ are arbitrary functions of the scalar field $\phi$ and its canonical kinetic term $2X\equiv - g^{\mu\nu}\phi_{\mu}\phi_{\nu}$.
The subscripts $\phi$ and $X$ denote partial derivatives ($G_{iX}\equiv\partial G_{i}/\partial X$), and $\epsilon^{\mu\nu\rho\sigma}$ is the antisymmetric Levi-Civita tensor.

As suggested by \cite{Langlois:2015cwa} and shown in \cite{Crisostomi:2016tcp}, this action as it is written can lead to an extra propagating degree of freedom.
However, this issue can be avoided by imposing the following condition \cite{BenAchour:2016fzp} ,
\begin{equation}
2XG_{5X}F_4 = -3F_5\left[G_4+4XG_{4X}+XG_{5\phi}\right]\,.
\end{equation}
This is visible in the context of beyond Horndeski when non-linear terms are considered. However, in this work we only consider linear perturbations and do not need to impose this condition in obtaining our results.

Starting from this action, one has to choose the functional form of the $G_i$ (and $F_i$), initial conditions for the scalar field, and get predictions solving for the background and the perturbations. 
However, this can be expensive and inefficient if the objective is to test the robustness of the standard cosmological model just by detecting deviations from it.
A more practical approach for this purpose is to use parametrized frameworks which were built to compress the information of a general class of models into few functions of time.
In particular we are going to use the framework introduced in \cite{Bellini:2014fua, Gleyzes:2014rba}, where it was shown that the expansion history of the universe and the linear perturbations can be described completely by six functions of time: one responsible for the background evolution -- $H(t)$ or $w(t)$ -- and the remaining affecting only the perturbations.
It is possible to show that for a general theory that can be described as a sub-class of beyond Horndeski these functions are completely independent of the background and each other, which means that they represent the minimal set of functions that can describe all models within this set of theories.
Here we describe their physical meaning and definitions in terms of the $G_i$ and $F_i$ functions:
\begin{itemize}
\item The kineticity, $\kin$, is a generalisation of the standard kinetic term of simple DE models.
On its own this parameter describes some of the simpler DE models, with the only modification to $\Lambda$CDM being the addition of an extra fluid (e.g.~quintessence).
It is defined as,
\begin{equation}
\begin{aligned}
\kin \Ms & = 
4 H \dot{\phi} X\left[3 G_{5X} + X\left(7 G_{5X X} - 120 F_5 + 2X \left(G_{5X X X} - 66 F_{5X} - 12 X F_{5X X}\right)\right)\right]\\
&+ \frac{2 X}{H^2}\left[K_{X} - 2 G_{3\phi} + + 2 X \left(K_{X X} - G_{3\phi X}\right)\right]\\
&+ \frac{12 \dot{\phi} X}{H}\left[G_{3X} - 3 G_{4\phi X} + X\left(G_{3X X} - 2 G_{4\phi X X}\right)\right]\\
&+ 12X \left[G_{4X} - G_{5\phi} + X\left( 8 G_{4X X} - 5 G_{5\phi X} + 2X\left(2 G_{4X X X} - G_{5\phi X X} \right)\right)\right]\\
&+ 12X\left[X\left(24 F_4 + 2X\left( 18 F_{4X} + 4 X  F_{4X X} \right)\right)\right]\,.
\end{aligned}
\end{equation}

\item The braiding, $\bra$, describes the mixing of the scalar field and metric kinetic terms.
It contributes to the kinetic energy of the scalar, partially sourced through its mixing with the metric.
In terms of $G_i$ and $F_i$,
\begin{equation}
\begin{aligned}
\bra \Ms &= 2 H \dot{\phi} X \left(3 G_{5X} + 2 X G_{5X X} - 60 X F_5 - 24 X^2 F_{5X}\right)\\
&+ \frac{2 \dot{\phi}}{H}\left[ - G_{4\phi} + X\left( G_{3X} - 2 G_{4\phi X}\right)\right]\\
&+ 8 X \left[G_{4X} - G_{5\phi} + X \left(2 G_{4X X} - G_{5\phi X} + 8 F_4  + 4 X F_{4X}\right)\right]\,.
\end{aligned}
\end{equation}

\item The effective Planck mass,
\begin{equation}
\Ms = 2 G_4 - 2 X\left[ 2 G_{4X} +  H \dot{\phi} G_{5X} - G_{5\phi} + 2 X \left( 2 F_4 - 6 H \dot{\phi} F_5\right)\right]\,,
\end{equation}
can be absorbed in the densities and pressures of matter and the scalar and be effectively hidden in the equations of motion.
But this is not the case for the Planck mass run rate,
\begin{equation}
\begin{aligned}
\run \Ms &\equiv\frac{{\rm d}\ln\Ms}{{\rm d}\ln a} = 
\frac{2 \dot{H} \dot{\phi} X}{H}\left(-G_{5X} + 12 X F_5\right) \\
& + 2 H \dot{\phi} X\left[3G_{5X} + 2X \left(G_{5X X} - 30 F_5 - 12 X F_{5X}\right)\right]\\
& - 2 \ddot{\phi} X \left[3 G_{5X} + 2X\left(G_{5X X} - 30 F_5 - 12 X F_{5X}\right)\right] \\
& - \frac{2 \dot{\phi} \ddot{\phi}}{H}\left[G_{4X} - G_{5\phi} + X \left(2 G_{4X X} - G_{5\phi X} + 8 F_4 + 4 X F_{4X}\right) \right]\\
& + \frac{2 \dot{\phi}}{H}\left[G_{4\phi} + X \left( G_{5\phi \phi} - 2 G_{4\phi X} - 4 X F_{4\phi}\right)\right]\\
& + 4X \left[ G_{4X} - G_{5\phi} + 2X \left(G_{4X X} - G_{5\phi X} + 4 F_4 + 2 X \left(F_{4X} + 3 F_{5\phi}\right)\right) \right]\,,
\end{aligned}
\end{equation}
which measures the variation in time of the Planck mass. This function is non-zero in models where in the action the scalar field couples directly to curvature, and it produces anisotropic stress in the gravitational potentials.

\item The tensor speed excess, $\ten$, signifies deviations of the speed of gravitational waves from the speed of light and has an effect on the scalar perturbations too.
As $\run$, it is present in non-minimally coupled theories and can be written as,
\begin{equation}\label{equ:tensor}
\ten \Ms = X\left[2 G_{5X} \left(2 H \dot{\phi} - \ddot{\phi}\right) +
4\left(G_{4X} - G_{5\phi} + 2 X \left(F_4 - 3 F_5 H \dot{\phi}\right)\right)\right]\,.
\end{equation}

\item $\beh$ is non-zero for beyond Horndeski theories.
It has been shown that even when ordinary mater is minimally coupled to the metric, in the theories beyond Horndeski, the higher order derivatives in the conformal transformation of the metric can lead to coupling between the sound speeds of dark energy and matter \cite{Gleyzes:2014dya, Gleyzes:2014qga, DAmico:2016ntq}.
This mixing between the scalar and matter can be measured by the $\beh$ parameter,
\begin{equation}
\beh \Ms = 8 X^2 \left(F_4 - 3 F_5 H \dot{\phi}\right)\,.
\end{equation}
\end{itemize}

Here, the dots denote derivatives with respect to proper time.
As a shorthand notation in the rest of this work we may refer to these functions as $\alpha_i$.

It is possible to derive the linear perturbation equations and an evolution equation for the perturbation of the scalar in term of these functions.
These equation, combined with the evolution equations for the matter species, provide the full system that one has to solve to obtain the linear matter and CMB power spectra and any other prediction within linear theory.
Given the constraining power of current datasets, any practical use of the $\alpha_i$ implies the choice of a parametric form to describe their time evolution.
There are many different parametrisations that can be used.
Although they cannot fully represent the beyond Horndeski class of theories, we have to make a choice in order to evolve the equations of motion in time.
For example one could choose to keep the $\alpha_i$ constant, split them in redshift bins, or evolve them as proportional to a power of the scale factor $a^{n}$ or the the fractional density of the DE/MG component $\Omega_s^n$, with $n=const$.
Here we pick the parametrisation proposed in \cite{Bellini:2014fua},
\begin{equation}
    \alpha_i = \Omega_s\hat{\alpha}_i\,,
    \label{equ:propto_omega}
\end{equation}
where $\hat{\alpha}_i$ are constants. We choose this time evolution because we are interested in DE/MG models that aim at explaining the late-time acceleration of the universe, so it is reasonable to assume that the $\alpha_i$ will become more important as the DE density grows and the universe starts to accelerate.
If all the $\hat{\alpha}_i$ turn out to be negligibly small, that would suggest that the accelerated expansion is mainly due to a cosmological constant.
Furthermore, this parametrisation provides more freedom than the constant or binned $\alpha_i$, but has a smaller parameter space to sample from than for $\alpha_i = \Omega_s^{n_i}\hat{\alpha}_i$. 
While the reasons we have invoked are practical, we note, however, that the study of specific beyond Horndeski models is in its infancy; specifically, there is, as yet, little understanding of the type of behaviour one might expect if constructing a model from the ground up (i.e. via an action). 
It would be important to explore the range of possible behaviours (as has been done for the case of the equation of state and the CLP parametrization in the case of dark energy and, to a lesser extent, as has been done for "normal" Horndeski models) -- only then can we be assured that such a parametrisation is representative of a model space.

\subsection{Background}

The Friedmann equations for a flat Friedmann-Robertson-Walker (FRW) metric read
\begin{equation}
    \begin{aligned}
    &3H^2=\trhom + \trhos\,,\\
    &2\dot{H}=-3H^2-\tpm-\tps\,,\label{equ:back}
    \end{aligned}
\end{equation}
where the subscript $m$ stands for all matter species and $s$ for DE/MG.
Here we have absorbed the Planck mass into the definitions of the densities and the pressures of matter and the scalar, e.g.~$\tilde{\rho}\equiv\rho/M_*^2$ (see \cite{Bellini:2014fua} for notation). As a consequence, the matter and DE densities are not conserved anymore
\begin{equation}
    \begin{aligned}
    &\dot{\tilde{\rho}}_{\rm m} + 3H(\trhom + \tpm) = -\run H\trhom\,,\\
    &\dot{\tilde{\rho}}_{\rm s} +3H(\trhos+\tps)=\run H\trhom\,.\label{equ:cons}
    \end{aligned}
\end{equation}
This can be interpreted as energy density being exchanged between matter and the scalar whenever $\Ms$ varies.
Thus in this framework we need to integrate the densities of all matter species appropriately according to the equations above.
With different definitions of DE/MG energy density and pressure we could have assumed the standard evolution for these quantities at the price of keeping track the value of $\Ms$.
Clearly the two approaches are equivalent and one uses whichever is more convenient. 
The expressions for $\trhos$ and $\tps$ in terms of the Horndeski $G_i$ and $F_i$ functions are given by,
\begin{equation}
    \begin{aligned}
         \Ms\trhos& \equiv - G_2 +2X\left(G_2 - G_{3\phi}\right) + 6\dot{\phi}H\left(X G_{3X} - G_{4\phi} - 2X G_{4\phi X}\right)\\
         &+ 12H^2X\left(G_{4X} + 2X G_{4X X} - G_{5\phi} - X G_{5\phi X} + 8X F_4 + 4 X^2 F_{4X}\right)\\
         &+ 4\dot{\phi}H^3 X\left(G_{5X} + X G_{5X X} - 24 X F_5 - 12 X^2 F_{5X}\right)\,,
    \end{aligned}
\end{equation}

\begin{equation}
    \begin{aligned}
        \Ms\tps&\equiv G_2 -2X\left(G_{3\phi} - 2G_{4\phi \phi}\right) + 4\dot{\phi}H\left(G_{4\phi} -2X G_{4\phi X} + X G_{5\phi \phi} - 4 X^2 F_{4\phi}\right)\\
        & -  \Ms\bra H\frac{\ddot{\phi}}{\dot{\phi}} 
        - 4 H^2 X^2 \left(G_{5\phi X} - 12X F_{5\phi}\right)
        + 2\dot{\phi} H^3 X \left(G_{5X} - 12F_5 \right)\,.
    \end{aligned}
\end{equation}

\subsection{Stability conditions}\label{sec:stability}

The price of introducing new degrees of freedom is that the resulting theory may be unstable on a given background.
Indeed, it is possible for specific choices of the $\alpha_i$ that the perturbations grow exponentially.
In this section we show the conditions to avoid two types of instabilities, i.e.~\textit{ghost} and \textit{gradient}.
\textit{Ghost instabilities} happen when we choose the wrong sign of the kinetic term of a d.o.f., while \textit{gradient instabilities} can occur when the sound speed of the fluid is imaginary.

After decoupling from auxiliary variables the quadratic action for scalar and tensor modes in beyond Horndeski with no matter reads \cite{DeFelice:2011bh,Gleyzes:2013ooa},
\begin{equation}
    S^{(2)}=\int d^4xa^3\left[Q_{\rm S}\left(\dot{\zeta}^2-\frac{\cs}{a^2}(\partial_i\zeta)\right) + Q_{\rm T}\left(\dot{h}_{ij}\,^2-\frac{c_{\rm T}^2}{a^2}(\partial_kh_{ij})^2\right)\right]\,,
\end{equation}
where $h_{ij}$ are the tensors modes (gravitational waves), and $\zeta$ the scalar.
To ensure that the propagating degrees of freedom are not ghost-like, we require that their kinetic terms are positive, i.e.
\begin{equation}
    \begin{aligned}
    &Q_{\rm S}\equiv\frac{2\Ms D}{(2-\bra)^2}>0\,,\qquad D\equiv\kin+\frac{3}{2}\bra^2\,,\\
    &Q_{\rm T}\equiv \frac{\Ms}{8}>0\,.
    \end{aligned}
\end{equation}
To avoid gradient instabilities we need to require that the speed of sound of the scalar and tensor degrees of freedom are positive.
This means that,
\begin{equation}
\begin{aligned}
    &c_{\rm s}^2 =\frac{(2-\bra)^2}{2D}\Biggl\{-c_{\rm T}^2 + \frac{4}{a\Ms}\frac{d}{dt}\left[\frac{a\Ms(1+\beh)}{H(2-\bra)}\right]\Biggr\} - \frac{\trhom+\tpm}{D H^2} (1+\beh)^2>0\,,\\
    &c_{\rm T}^2 = 1+ \ten > 0\,.
    \label{equ:speed}
\end{aligned}
\end{equation}
Note that the only condition affected by the beyond Horndeski terms is the gradient condition for the scalar sector, while the remaining are the same as for Horndeski models \cite{Kobayashi:2011nu, Bellini:2014fua}.

There have been studies of the impact that stability conditions have on the constraints obtained for parametrised models \cite{Salvatelli:2016mgy,Perenon:2016blf,Peirone:2017lgi} and \texttt{hi\_class} gives you the freedom to relax or disable the restrictions imposed by these conditions.
From a brief inspection of the results that we obtain with and without these restrictions we find that the combinations of $\alpha_i$ that get rejected by the stability conditions are also in regions of extremely low likelihood; if the tests are disabled they lead to exponential growth of the perturbations (and hence are an extremely bad fit to data).
While it is safe to ignore the restrictions,  we have decided to keep them in our analysis, as this speeds up sampling significantly.

\subsection{Recent implications from GW170817 and GRB 170817A}\label{sec:gw17}
After the detection of an electromagnetic counterpart (GRB 170817A) to the gravitational wave signal (GW170817) from a binary neutron star merger \cite{PhysRevLett.119.161101,2041-8205-848-2-L14,2041-8205-848-2-L15}, it has been shown that the speed of gravitational waves (GW) has to be incredibly close to the speed of light and thus $\left|\ten\right|<10^{-15}$ (we note that
Current constraints on $\ten$ using cosmological data are ${\cal O}(1)$).
This implies that, as pointed out in \cite{Baker:2017hug, Ezquiaga:2017ekz,Creminelli:2017sry,Sakstein:2017xjx}, this result imposes strict constraints on general scalar-tensor and vector-tensor theories that are much tighter than current cosmological constraints and therefore it is safe to take 
\begin{equation}
	\ten=0\,,\,\,\ c_{\rm T}^2=1\,. 
\end{equation}
It has been suggested that this  constraint can be avoided for some scalar-tensor theories if one takes into account the dynamics of the scalar field directly, when it is not coupled directly to matter \cite{Copeland:2018yuh}. 

Furthermore, in Ref.~\cite{Creminelli:2018xsv} the authors claim that any model with $\beh\neq 0$ predicts a copious decay of GW into DE scalar fluctuations. Given that that we do observe GW, suggests that we must rule out any beyond Horndeski model. The only way to avoid decay of GW and have $\beh\neq 0$ is to impose the propagation speed of scalar perturbations, $\cs$, to be equal to the speed of light. While the authors mention that there could be power-law divergent terms even in this case, here our objective is to give a purely phenomenological idea of the behaviour of beyond Horndeski theories. Our main analysis is then performed letting $\beh$ free to vary and without any additional condition on $\cs$ (beside the stability condition mentioned in the previous section). However, for completeness we compare this case with the one where we fix $\cs=1$ and show the results in Section~\ref{sec:results}.

It is possible to get models with $\cs = 1$ by expressing one of the $\alpha_i$ in terms of the others. In this paper we choose to fix $\kin$, which obeys this relation
\begin{equation}
\kin = -\frac{3}{2}\bra^2 - \frac{1}{2}(2-\bra)^2\Biggl\{c_{\rm T}^2 - \frac{4}{a\Ms}\frac{d}{dt}\left[\frac{a\Ms(1+\beh)}{H(2-\bra)}\right]\Biggr\} - \frac{\trhom+\tpm}{H^2} (1+\beh)^2\,.
\label{equ:kin_fix}
\end{equation}
This ensures that the speed of sound of the scalar is unity for all the models we consider at all times.
As shown in \cite{Bellini:2015xja}, the value of $\kin$ is generally unconstrained by data and varying this parameter has negligible effect on the background.
In Section~\ref{sec:phenomenology}, we can see that $\kin$ is not present in the equations of motion in the quasi-static approximation, Eqs.~(\ref{equ:QSphi}) and (\ref{equ:QSpsi}), so it also has no effect on the observables we are looking at in this limit.
In Section~\ref{sec:constr} we present the results for $\kin$ as given by Eq.~(\ref{equ:kin_fix}) and compare those to the case where $\kin = 1$.

\subsection{Perturbations}\label{sec:perts}
We assume that the universe is spatially flat and is well described by small perturbations to the Friedmann-Robertson-Walker (FRW) metric. 
Taking into account only scalar perturbations, the line element in Newtonian gauge (with the notation of \cite{Ma:1995ey}) reads,
\begin{equation}
    ds^2=-\left(1+2\Psi\right)dt^2+a^2(t)\left(1-2\Phi\right)d\mathbf{x}^2\,.
\end{equation}
Following \cite{Bellini:2014fua} we redefine the perturbations of the scalar field as
\begin{equation}
    v_X\equiv-\frac{\delta\phi}{\dot{\phi}}\,.
\end{equation}

Here we present the linear perturbation equations for the Fourier components of the action given in Eq.~(\ref{equ:action}) in Newtonian gauge and physical time. 
In Appendix \ref{app:hi-class} we present these in synchronous gauge, as implemented in \texttt{hi\_class}.
These equations have been derived previously in \cite{Zumalacarregui:2013pma, Gleyzes:2014dya}.\\
The modified Einstein (00) equations has the form,
\begin{equation}
\begin{aligned}
    &3\left(2 - \bra\right) H \dot{\Phi} + \left(6 - \kin - 6 \bra\right) H^2 \Psi + \left(1 + \beh\right)\frac{2 k^2}{a^2}\Phi \\
    & - \left(\kin + 3\bra\right)H^2\dot{v_X} - H\left[\left(\bra + 2 \beh\right)\frac{k^2}{a^2} - 3 \bra \dot{H} + 3\left( 2\dot{H} + \trhom + \tpm\right) \right]v_X  = - \trhom \delta_{\text{m}}\,.
\end{aligned}\label{equ:00}
\end{equation}
\noindent
the Einstein (0i),
\begin{equation}
2 \dot{\Phi} + \left(2 -  \bra \right) H \Psi - \bra H \dot{v_X} - \left(2 \dot{H} + \trhom + \tpm\right)v_X  = - \left(\trhom + \tpm\right)v_{\rm m}\,,\label{equ:0i}
\end{equation}
\noindent
the traceless part of the Einstein (ij) equation,
\begin{equation}
 \left(1 + \beh\right) \Psi - \left(1 +  \ten\right) \Phi + \beh \dot{v_X} - \left( \run - \ten \right)H v_X = - \frac{3}{2}\left(\trhom + \tpm\right) \sigma_m\,,\label{equ:tless}
\end{equation}
\noindent
and the trace part,
\begin{equation}
\begin{aligned}
    2 \ddot{\Phi} &-  \bra H \ddot{v_X} + 2\left(3 + \run \right) H \dot{\Phi} + \left(2 - \bra\right)H\dot{\Psi}\\
    &+ \left[\left(2 - \bra\right)\left(3 + \run\right)H^2 -  \left(\bra H\right)\dot{}\, - \left(2\dot{H} + \trhom + \tpm \right)\right]\Psi \\
    &- \left[\bra\left(3 + \kin \right) H^2 + \left(\bra H\right)\dot{}\, +  \left(2\dot{H} + \trhom + \tpm \right)\right]\dot{v_X} \\
    &- \left[2 \ddot{H} + 2\left(3 + \run\right) \dot{H} H +  \dot{\tilde{p}}_{\rm m} + \run H \tpm\right] v_X  = - \frac{k^2}{a^2}\sigma_m \left(\trhom + \tpm\right) + \delta \tpm\,,
\end{aligned}\label{equ:trace}
\end{equation}
\noindent
and the equation of motion for the scalar velocity potential $v_X$,
\begin{equation}
\begin{aligned}
    3 \bra \ddot{\Phi} H& + \kin H^2 \ddot{v_X} + 
    3\left[\bra\left(3 + \run\right) H^2 + \left(\bra H\right)\dot{}\, - \left(2\dot{H} + \trhom + \tpm\right)\right] \dot{\Phi} + \left(\kin + 3 \bra \right)H^2\dot{\Psi} \\
    & -  2\beh\frac{k^2}{a^2}\dot{\Phi} - 2 \Bigl[\dot{\beh} + \left(\beh \left(1 + \run\right)+ \run - \ten\right) H \Bigr]\frac{k^2}{a^2}\Phi  - \left(\bra + 2 \beh \right)H \frac{k^2}{a^2}\Psi\\
    &+ \left[\left(2\kin + 9 \bra\right) \dot{H} + \left(\dot{\kin} + 3 \dot{\bra}\right) H + \left(3 + \run\right)\left(3\bra + \kin\right)H^2 -3\left( 2\dot{H} + \trhom +\tpm\right)\right]H\Psi\\
    &-3 \Bigl\{3\dot{H}\left[\left(\bra H\right)\dot{}\, - \left(2\dot{H} + \trhom + \tpm \right)\right] + 3 \bra \left[\ddot{H} +  \left(3 + \run\right)\dot{H}H\right]H\Bigr\} v_X\\
    &+ \Bigl\{\left[\left(\bra +2\beh\right)H\right]\dot{}\, + \left(\bra  + 2 \beh\right)\left(1 + \run \right)H^2 + 2 \left(\run - \ten\right)H^2 - 2 \left(\dot{H} + \trhom + \tpm\right)\Bigr\} \frac{k^2}{a^2}v_X\\
    &+\left[2 \kin \dot{H} + \dot{\kin} H + \kin\left(3 + \run\right)\right] H\dot{v_X} = 0\,.
\end{aligned}\label{equ:pert}
\end{equation}
These equations, combined with the evolution equations of the matter perturbations, $\delta_m$, $v_m$, $\delta \tpm$, and $\sigma_m$, form the complete set of equations that one has to solve to follow the linear dynamics of beyond Horndeski theories.
We have implemented these equations into \texttt{hi\_class} \cite{Zumalacarregui:2016pph} and used the solutions to study the phenomenology of this class of models and constrain its parameters with current CMB, large-scale structure data in the following two sections.

\section{Phenomenology of beyond Horndeski theories}\label{sec:phenomenology}
The purpose of this work is to provide constraints on parametrized beyond Horndeski models, i.e.~using $\beh$.
In this section we briefly discuss the observables we use to calculate our parameter confidence regions, namely the matter power spectrum and the CMB temperature, polarisation and lensing power spectra.
We present the results obtained with \texttt{hi\_class} for the evolution of these quantities for different values of $\beh$ and derive the analytic expressions for these in the quasi-static approximation and observe what effects $\beh$ has on them.

\subsection{The quasi-static approximation}\label{sec:qs}
The key assumption of the quasi-static approximation (QSA) is,
\begin{equation}
|\dot{X}|\lesssim H|X|\,,
\end{equation}
where $X$ stands for any metric or scalar perturbation.
This assumption states that the time evolution of all parameters and perturbations of the metric and the scalar is comparable to the evolution of the Hubble parameter.
However, this can be only true on sub-horizon scales, where space-derivatives become important
\begin{equation}
\frac{k^2}{a^2}|X|\gg H^2|X|\,.
\end{equation}

We can apply this limit to the Hamiltonian constraint equation,  Eq.~(\ref{equ:00}), and the equation of motion for the scalar perturbation, Eq.~(\ref{equ:pert}),
\begin{equation}
    \frac{2 k^2}{a^2}\left(1 + \beh\right)\Phi - \frac{k^2}{a^2}\left(\bra + 2 \beh\right)H v_X = - \trhom \delta_{\rm m}\,,
\end{equation}

\begin{equation}
\begin{aligned}
2& \beh \frac{k^2}{a^2}\dot{\Phi} +  \left(\bra + 2 \beh\right)\frac{k^2}{a^2} H \Psi + 2\left[\dot{\beh} + \Bigl(\beh (1 + \run) + \run - \ten \right)H\Bigr]\frac{k^2}{a^2} \Phi \\
&+ \Bigl\{\left[\left(\bra +2\beh\right)H\right]\dot{}\, + \left[\left(\bra  + 2 \beh\right)\left(1 + \run \right) + 2 \left(\run - \ten\right)\right]H^2 - \left(2\dot{H} + \trhom + \tpm\right)\Bigr\} \frac{k^2}{a^2}v_X = 0\,.
\end{aligned}
\end{equation}
In this approximation, we neglect any contribution from the matter anisotropic stress, $\sigma_m$, and pressure, $\tpm$.
The reason is that the QSA can give predictions only at late-times, where the dominant matter component is DM, which is modelled as a pressureless perfect fluid.

Combining these with the rest of the constraint equations Eqs.~(\ref{equ:0i} - \ref{equ:trace}) and their time derivatives, we get the constraint equations for the perturbations $\Phi$ and $\Psi$ in this limit,
\begin{align}
    &\frac{k^2}{a^2}\Phi = -\frac{3}{2}H^2\Omega_{\rm m}\mu_\Phi\delta_{\rm m} + \beh \lambda_\Phi H\frac{k^2}{a^2}v_{\rm m}\,,\label{equ:QSphi}\\
    &\frac{k^2}{a^2}\Psi = -\frac{3}{2}H^2\Omega_{\rm m}\mu_\Psi\delta_{\rm m} + \beh \lambda_\Psi H\frac{k^2}{a^2}v_{\rm m}\,,\label{equ:QSpsi}
\end{align}
where
\begin{align*}
    \mu_\Phi& = \frac{1}{1+\beh}\left(1+\frac{\bra+2\beh}{\csn}B\right)\,,\\
    \lambda_\Phi& = -\frac{1}{2}\frac{(\bra+2\beh)}{\csn H^2}\trhom\,,\\
    \mu_\Psi& = \frac{1}{(1+\beh^2A)(1+\beh)^2}\left[c_{\rm T}^2 + 2\left(B+\frac{\dot{\beh}}{H}\right)\frac{B}{\csn}+\beh(1+\beh)a\Ms\frac{d}{dt}\left(\frac{2B}{aH\Ms\csn}\right)\right]\,,\\
    \lambda_\Psi& =\frac{\beh\dot{A}+2\dot{\beh}A}{1+\beh^2A} \,,
\end{align*}
\begin{equation*}
A\equiv\frac{2\lambda_\Phi}{(\bra+2\beh)H}\,,\qquad
B\equiv\frac{1}{2}\left(\bra +\bra\ten + 2\ten \kin - 2\kin\beh - 2\beh - \frac{2\dot{\beh}}{H}\right)\,. 
\end{equation*}
\begin{figure}
    \centering
    \begin{subfigure}[t]{0.48\textwidth}
        \includegraphics[width=\textwidth]{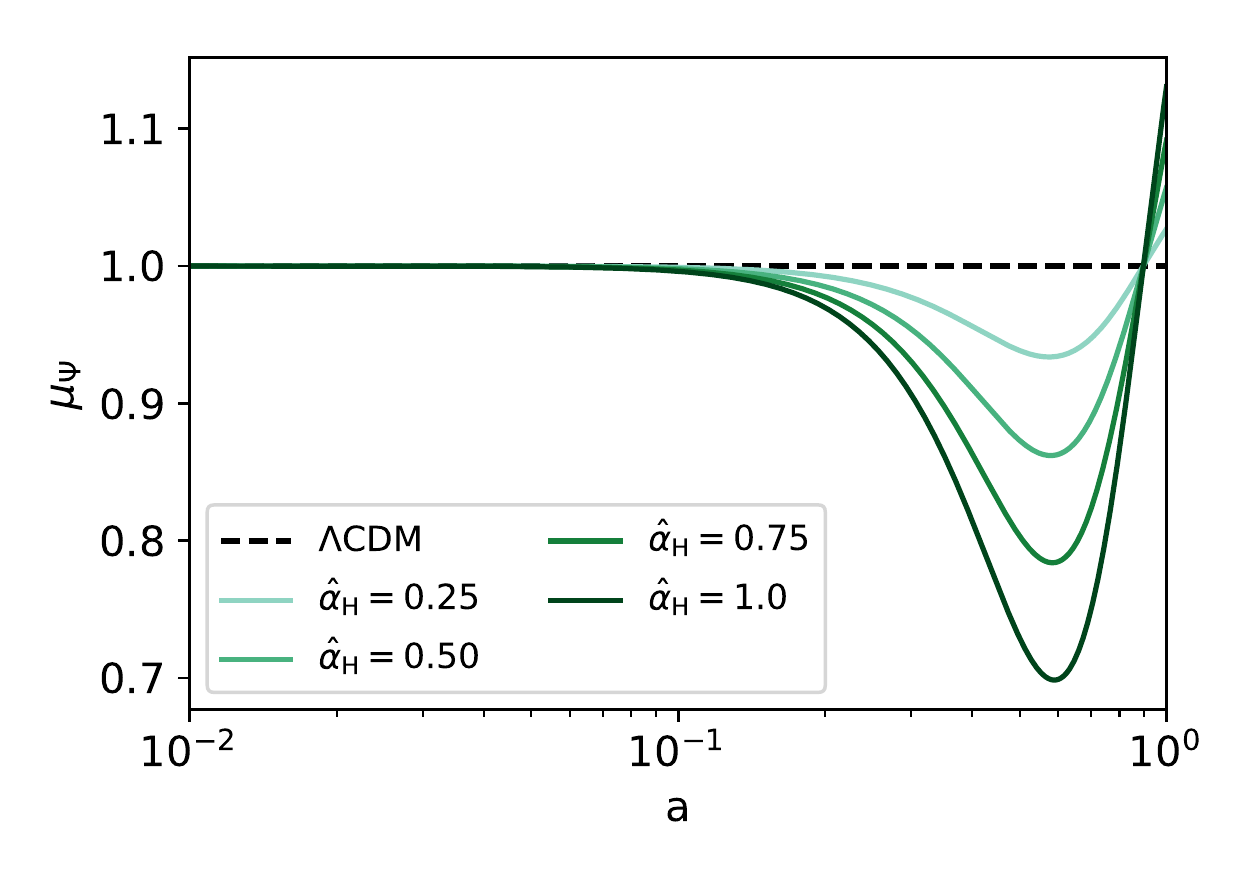}
    \label{fig:mu-psi}
    \end{subfigure}
    \,
    \begin{subfigure}[t]{0.48\textwidth}
        \includegraphics[width=\textwidth]{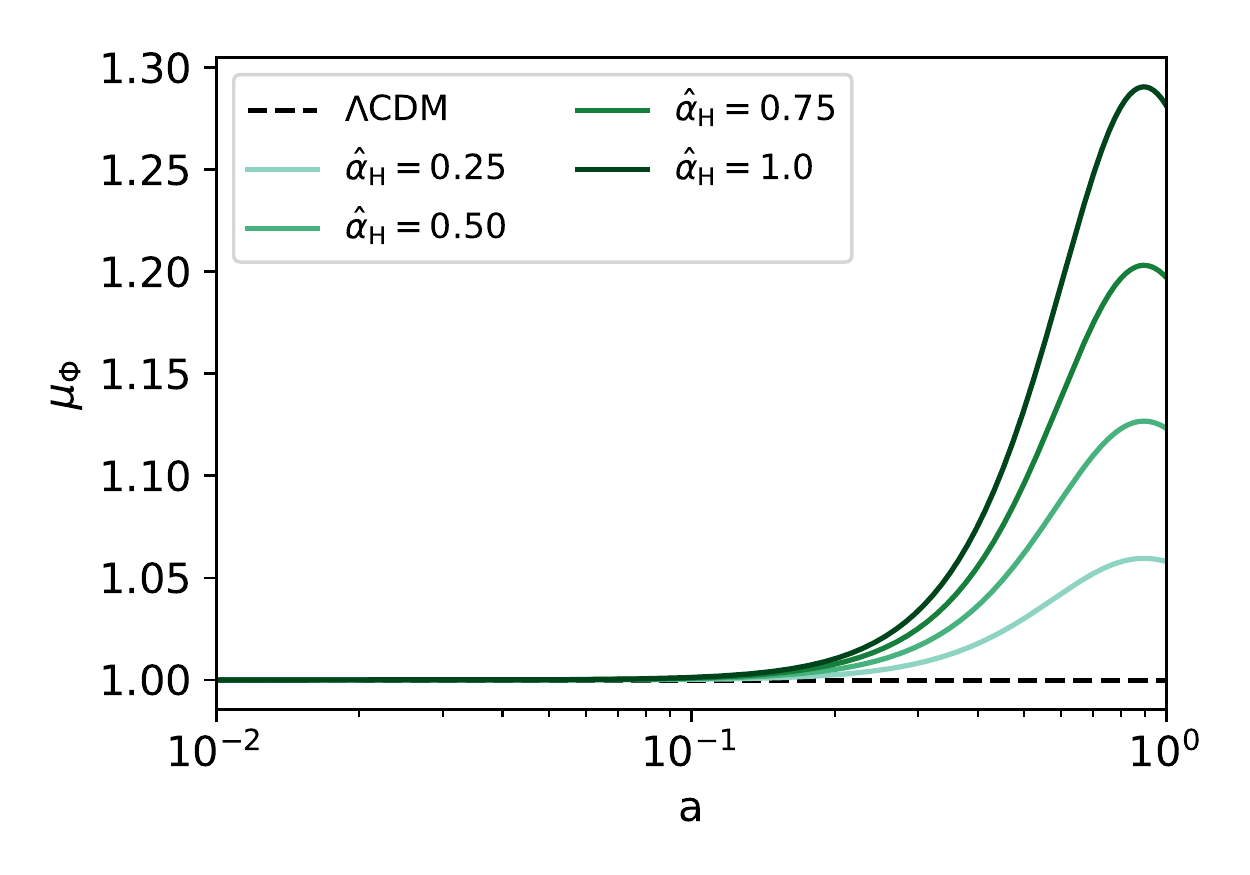}
        \label{fig:lsmbda-psi}
    \end{subfigure}
    \caption{The evolution of $\mu_\Psi$ (left) and $\mu_\Phi$ (right) as a function of scale factor.
    We have fixed $\bra=\run=\ten=0$, the background cosmological parameters to their estimated values from Planck \cite{Ade:2015xua} and chosen background expansion history consistent with  $\Lambda$CDM.}
    \label{fig:mus}
\end{figure}
These can be combined with the Euler and continuity equations for DM, i.e. 
\begin{equation}
\dot{\delta}_{\rm m}=\frac{k^2}{a^2}v_{\rm m}\,,\qquad\dot{v}_{\rm m}=-\Psi\,.
\label{equ:euler}
\end{equation}
Note that the notation we have used here is different from the commonly used effective gravitational constant, $\mu\equiv G_{\rm eff}/G$ and gravitational slip parameter, $\gamma\equiv\Phi/\Psi$.
In beyond Horndeski we have additional terms involving matter velocity that can not be neglected on sub-horizon scales, i.e.~the ones proportional to $\lambda_\Phi$ and $\lambda_\Psi$.
To avoid confusion we chose to use $\mu_\Phi$ and $\mu_\Psi$ instead.
In Horndeski ($\beh=0$), we would be able to relate the variables used here to the usual $\mu$ and $\gamma$ as
\begin{equation}
	\mu_\phi\equiv\mu\,\qquad\mu_\Psi\equiv\mu/\gamma\,.
\end{equation}
We know that in the $\Lambda$CDM model these are both equal to one.
So the strength of gravity in other models can be expressed as a the gravitational constant in $\Lambda$CDM plus a small correction \cite{Baker:2013hia}, 
\begin{equation}
    \mu = 1 + \delta\mu = -\frac{2k^2\Phi}{3H^2\Omega_{\rm m}\delta_{\rm m}}\,.
\end{equation}
However, here due to the presence of the beyond Horndeski parameter, $\beh$, this is not as straightforward. 
The effect of this additional term is non-negligible and we have taken it into account in all the calculations here.
To show the significance of these terms we look at the simplified case where $\bra=\kin=\ten=0$ and $\beh=\hbeh\,\Omega_{\rm DE}$, i.e.~a model with a standard kinetic term $\kin$ plus beyond Horndeski. The functions $\lambda_\Phi$ and $\mu_\Phi$ simplify to
\begin{equation}
    \lambda_\Phi= -\frac{3\Omega_{\rm m}}{2-3(\beh-1)\Omega_{\rm m}} \,,\qquad \mu_\Phi= 1 - \beh\lambda_\Phi\,,
    \label{equ:la-mu_phi}
\end{equation}
and for $\lambda_\Psi$ and $\mu_\Psi$, we have
\begin{equation}
    \lambda_\Psi=-\frac{9\Omega_{\rm m}\left(2-4\Omega_{\rm m}-3\Omega_{\rm m}^2\right)}{(2+3\Omega_{\rm m})\left[2 + 3(1-\beh)\Omega_{\rm m}\right]}\,,\qquad \mu_\Psi=1 - \beh\lambda_\Psi\,.
    \label{equ:la-mu_psi}
\end{equation}
$\mu_\Phi$ and $\mu_\Psi$ are plotted on Figure~\ref{fig:mus} as functions of the scale factor, $a$. 
In order to isolate its contribution, all $\alpha_i$ except for $\beh$ have been set to zero.
Given that we chose the evolution of $\beh$ to be proportional to $\Omega_{\rm DE}$, we see that the the deviations from $\mu_\Phi = \mu_\Psi = 1$ are very small before the onset of DE. 
On the left panel we have the evolution of $\mu_\Psi = 1 - \beh\,\lambda_\Psi$, which begins to decrease below one as $\beh$ becomes non negligible at late times.
On the right is $\mu_\Phi = 1 - \beh\,\lambda_\Phi$. At late times $\beh\,\lambda_\Phi$ becomes negative and enhances $\mu_\Phi$.
\begin{figure}
    \centering
    \includegraphics[width=0.6\textwidth]{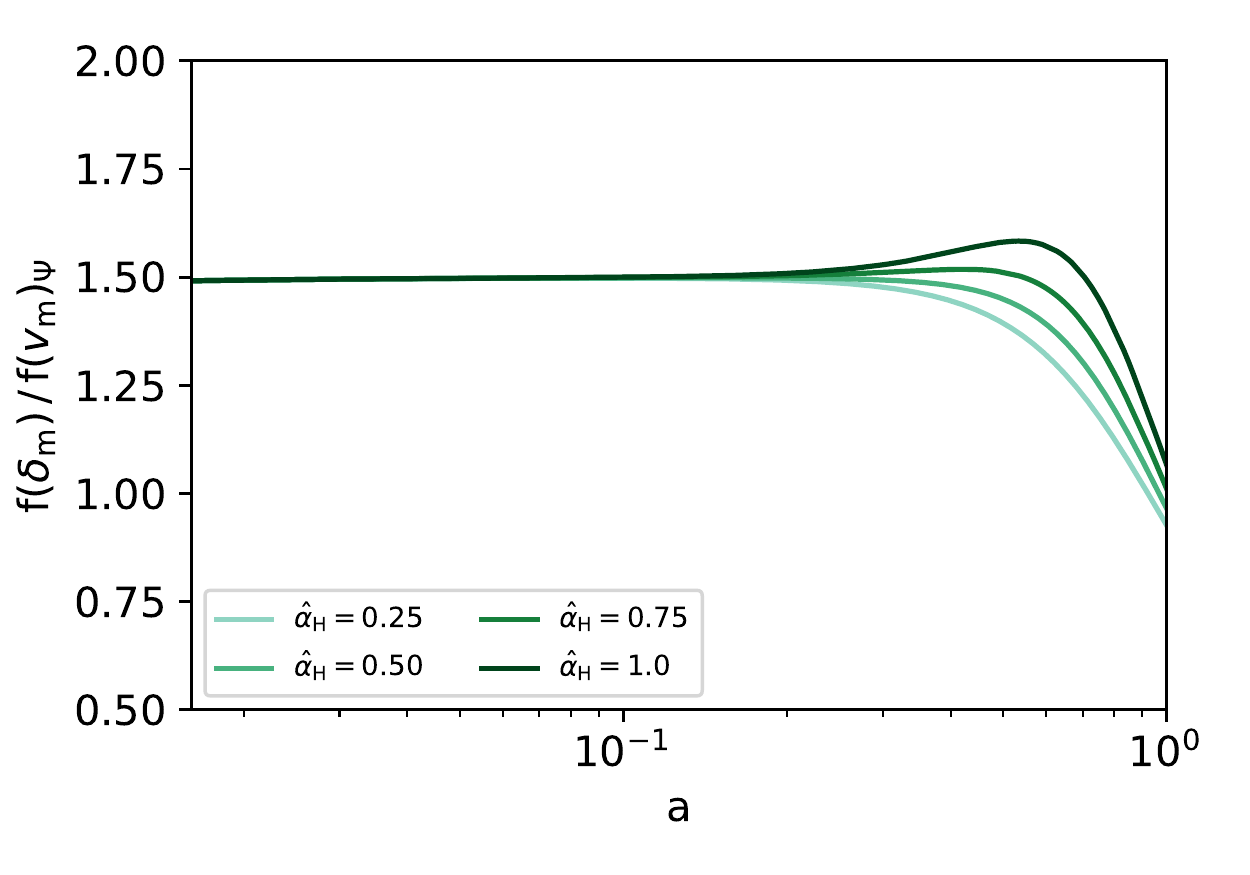}
    \caption{The ratio between the two terms contributing to the deviation from GR in the QSA equations, Eqs.~(\ref{equ:QSpsi} and \ref{equ:QSpsi}) for 4 different values of $\hbeh$ as a function of scale factor. On the $y$-axis we have abbreviated  ${\rm f}(\delta_{\rm m})\equiv\frac{3}{2}H^2\Omega_{\rm m}\beh \lambda_\Phi \delta_{\rm m}$ and ${\rm f}(v_{\rm m})\equiv\beh \lambda_\Phi H\frac{k^2}{a^2}v_{\rm m}$.}
    \label{fig:rats}
\end{figure}

In this case we can write $\delta\mu_\Phi\equiv-\beh \lambda_\Phi$ and so the modifications to GR coming from the two terms on the RHS of Eq.~(\ref{equ:QSphi}) are  $\frac{3}{2}H^2\Omega_{\rm m}\beh \lambda_\Phi \delta_{\rm m}$ and $\beh \lambda_\Phi H\frac{k^2}{a^2}v_{\rm m}$, and equivalently for Eq.~(\ref{equ:QSpsi}).
The ratio of these two contributions is plotted on Figure~\ref{fig:rats} for four different values of $\hbeh$.
We can see that the these are of the same order and hence have roughly equal contributions to the equations. 
This means that the $v_{\rm m}$ term cannot be neglected and must be taken into account in the definition of the QS approximation equations.
It is possible to redefine the effective gravitational constant and hence the constraint equations above using the common notation and taking into account the additional terms we have here.
As usual, we define the growth factor as the logarithmic derivative of the matter overdensity
\begin{equation}\label{equ:growth-def}
    f = \frac{d{\rm ln}\delta_{\rm m}}{d{\rm ln}a}\,.
\end{equation}
Using,
\begin{equation}
\begin{aligned}
    \frac{k^2}{a^2}v_{\rm m} = \dot{\delta}_{\rm m} = f H\delta_{\rm m}\,,
\end{aligned}
\end{equation}
we can re-write the constraint equation for $\Phi$ in its usual form
\begin{equation}
\frac{k^2}{a^2}\Phi = -\frac{3}{2}H^2\Omega_{\rm m}\mu\delta_{\rm m}\,,
\end{equation}
where
\begin{equation}
    \mu = \mu_\Phi - \frac{2}{3\Omega_{\rm m}}\beh\lambda_\Phi f\,.
\end{equation}
We show in Section~\ref{sec:growth} that the approximate solution for the growth factor in the quasi-static limit is very close to the exact solution, which makes this a good expression for the effective gravitational constant.

\subsection{Matter power spectrum}\label{sec:mpk}
\begin{figure}
    \centering
    \begin{subfigure}[t]{0.48\textwidth}
        \includegraphics[width=\textwidth]{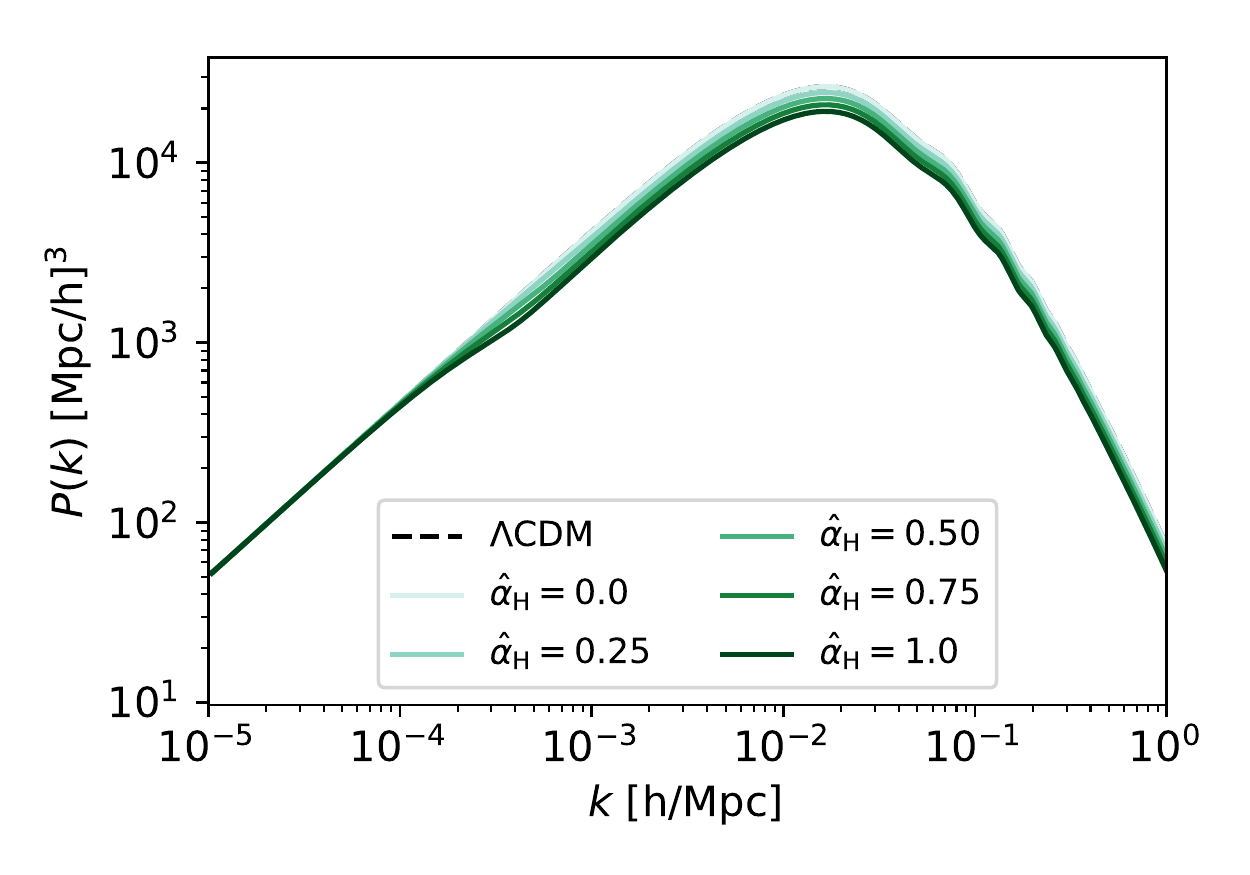}
    \end{subfigure}
    \,
        \begin{subfigure}[t]{0.48\textwidth}
        \includegraphics[width=\textwidth]{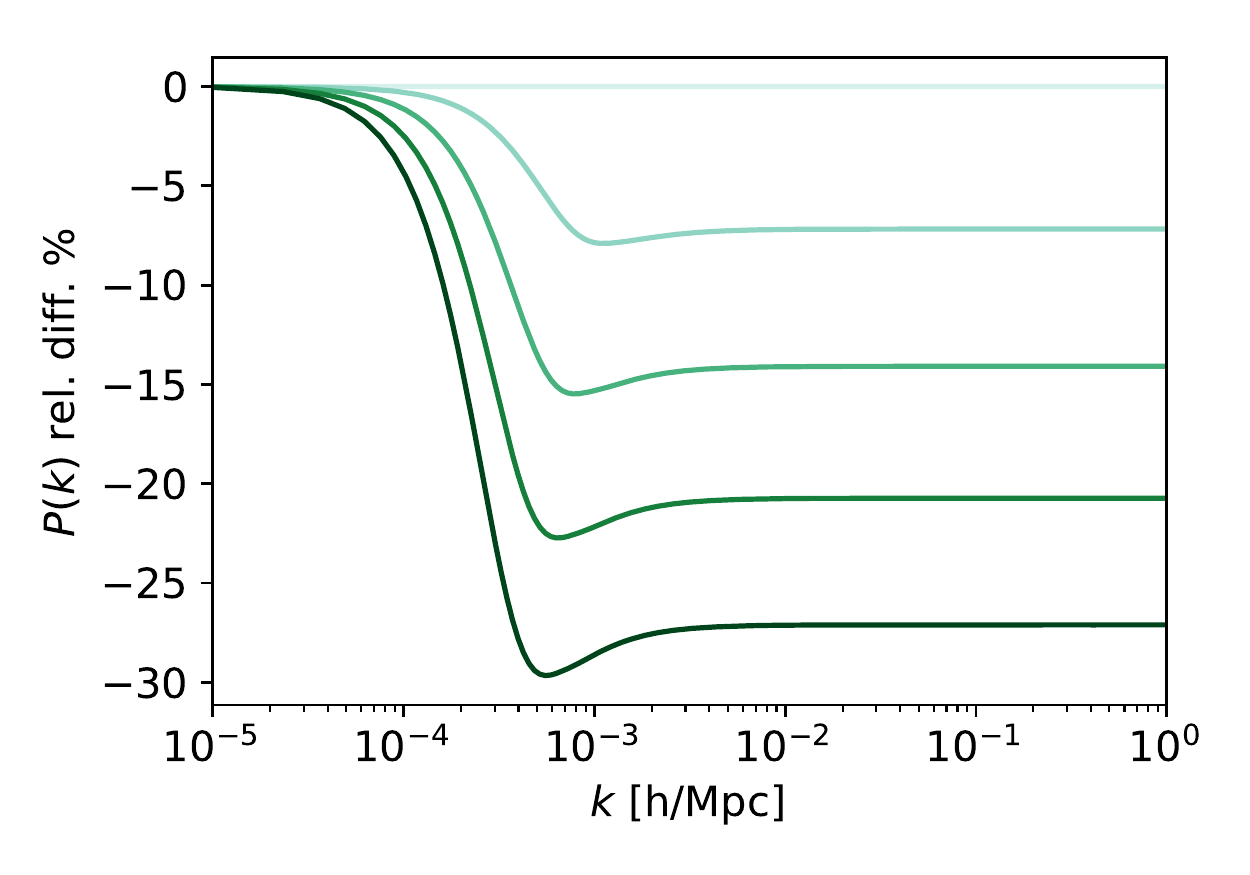}
    \end{subfigure}
    \caption{The effects of $\hbeh$ on the matter power spectrum at $z=0$. On the right graph we show the relative difference between the models we are considering and a fiducial $\Lambda$CDM model. For each line the background cosmological parameters are fixed to their estimated values from Planck \cite{Ade:2015xua}.
    For the background expansion history we assume a $\Lambda$CDM parametrisation (or $w_{\rm DE} = -1$) and for the $\alpha_i$ we choose the evolution proposed in \cite{Bellini:2014fua}. Here we fix $\hkin = 1$, $\hbra=\hrun=\hten=0$ and vary $\hbeh$.}
    \label{fig:mpk}
\end{figure}

Combining Eqs.~(\ref{equ:QSpsi}) and (\ref{equ:euler}), we can derive the evolution equation for the matter overdensity, $\delta_{\rm m}$, in the QS limit
\begin{equation}\label{eq:matter_qs}
    \ddot{\delta}_{\rm m} + (2+\beh\lambda_\Psi)H\dot{\delta}_{\rm m}-\frac{3}{2}H^2\Omega_{\rm m}\mu_{\Psi}\delta_{\rm m}=0\,.
\end{equation}
As we have seen above the presence of $\beh$ decreases the effective gravitational constant $\mu_\Psi$ by $\beh\lambda_\Psi$, which in our parametrisation becomes important only at late times (see Figure~\ref{fig:mus}).
Then, increasing the value of $\hbeh$, we expect a suppression of the matter power spectrum at small scales.
This has two contributions acting in the same direction: (i) a decreased value of the effective gravitational constant $\mu_\Psi$, and (ii) an enhanced friction term proportional to $\beh$.
In the matter-dominated era, where $a\propto t^{2/3}$, $H=2/(3t)$ and $\beh\lambda_\Psi\ll 1$, an approximate solution to Eq.~(\ref{eq:matter_qs}) up to linear order in $\beh$ is given by
\begin{equation}
    \delta_{\rm m}\propto t^{2/3(1-\beh\lambda_\Psi)}\,,
\end{equation}
where in the derivation we assumed constant coefficients. 
This confirms that, during this period, while $\beh\lambda_\Psi>0$, the growth of overdensities and hence the matter power spectrum have to be suppressed.

In Figure~\ref{fig:mpk} we show the matter power spectrum as a function of $k$ at redshift $z=0$ as outputted from \texttt{hi\_class}. 
We have set the other parameters, $\hkin=1$, $\hbra=\hrun=\hten=0$, in order to observe the effect of $\hbeh$ alone.
We set the background cosmological parameters to their best fit values form Planck 2015 \cite{Ade:2015xua} and assume $\Lambda$CDM for the background expansion history.
In this case only positive values of $\hbeh$ are allowed, as negative $\hbeh$ would cause a gradient instability.
Assuming $\Lambda$CDM background expansion we fix $w_s=p_s/\rho_s=-1$, so $p_s+\rho_s=0$ and $\dot{\rho}_s = 0$, and given that we chose $\beh=\Omega_s\hbeh$, the gradient instability, Eq.~(\ref{equ:speed}) simplifies to,
\begin{equation}
	\left[\left(\rho_m+p_m\right)\left(1-\beh\right) + 2H^2\right]\beh>0\,,
\end{equation}
which is only satisfied when
\begin{equation}
	\beh>0 \qquad\text{and}\qquad \beh < 1 + \frac{2H^2}{\rho_m+p_m}\,.
\end{equation}
For the time evolution of the $\alpha_i$ we assumed parametrisation in Eq.~(\ref{equ:propto_omega}).
On the left panel of Figure~\ref{fig:mpk} we show the matter power spectrum generated by \texttt{hi\_class} for five different values of $\hbeh$ and on the right the relative difference between those and our fiducial $\Lambda$CDM.
This is consistent with the results of \cite{DAmico:2016ntq}.
As expected and discussed in detail in this section, $\hbeh$ has a damping effect on the matter power spectrum and the deviation from $\Lambda$CDM on small scales indeed seems proportional to $\hbeh$.

\subsection{The growth rate of structure}\label{sec:growth}
\begin{figure}
    \centering
    \begin{subfigure}[t]{0.48\textwidth}
        \includegraphics[width=\textwidth]{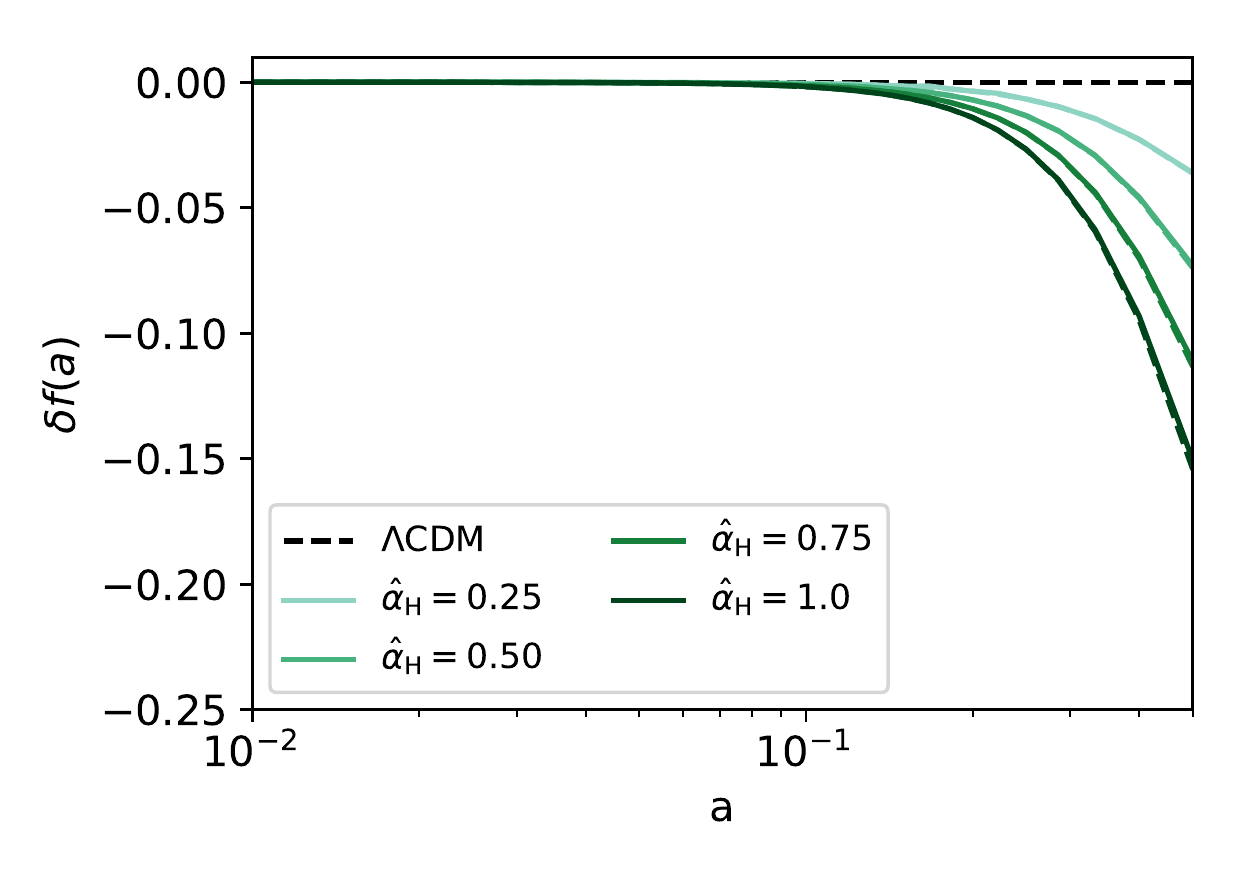}
    \end{subfigure}
    \,
    \begin{subfigure}[t]{0.48\textwidth}
        \includegraphics[width=\textwidth]{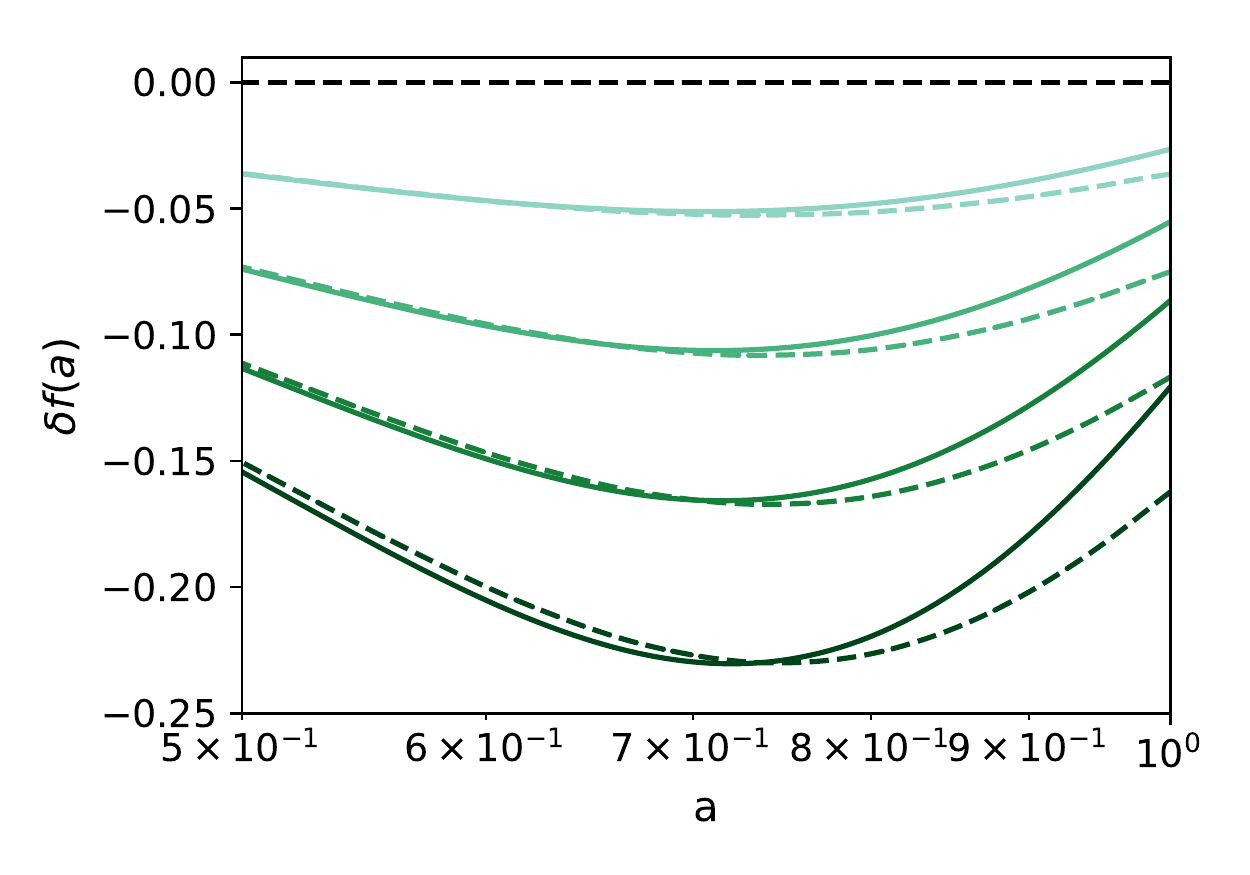}
    \end{subfigure}
    \caption{Relative difference of the growth rate, $f$, between 4 models with different $\hbeh$ and $\Lambda$CDM. On the left we have $10^{-2}<a<0.5$ and on the right, $0.5<a<1$. Solid lines are the results obtained with \texttt{hi\_class} and the dashed lines are the approximation, Eq.~(\ref{equ:deltaf}).}
    \label{fig:growth}
\end{figure} 
Using the definition of the growth rate of structure, Eq.~(\ref{equ:growth-def}), and the evolution equation for the matter overdensity, Eq.~(\ref{eq:matter_qs}), we have,
\begin{equation}
    f^{\prime} + f^2 + \left(2 + \beh\lambda_\Psi + \frac{\dot{H}}{H^2}\right)f - \frac{3}{2}\Omega_{\rm m}\mu_\Psi = 0\,,
    \label{equ:growth}
\end{equation}
where primes denote a derivative with respect to $\ln a$.
We can find an approximate solution to this equation in the matter-dominated era, making the following assumptions.
The growth rate of structure can be expressed as the growth rate of matter in matter domination plus a small correction that represents the DE/MG component
\begin{equation}
    f = f_{\rm m} + \delta f\,.
\end{equation}
Similarly, as argued above, we can express the effective gravitational constant $\mu_\Psi$, as its standard value, i.e.~$\mu_\Psi=1$, plus a small correction, $\delta\mu_{\Psi}\equiv-\beh\lambda_\Psi$.
Expanding the growth equation Eq.~(\ref{equ:growth}) to first order in the perturbation parameters, we get,
\begin{flalign}
    &0^{\rm th}\, \text{order}\hspace{2 cm}    f^{\prime}_{\rm m} + f_{\rm m}^2 + \left(2 + \frac{\dot{H}}{H^2}\right)f_{\rm m} - \frac{3}{2}\Omega_{\rm m} = 0\,,\\
    &1^{\rm st}\, \text{order}\hspace{2 cm}    \delta f^{\prime} + 2f_{\rm m}\delta f + -\delta\mu_{\Psi} f_{\rm m} + \left(2 + \frac{\dot{H}}{H^2}\right)\delta f - \frac{3}{2}\Omega_{\rm m}\delta\mu_{\Psi} = 0\,.&&
\end{flalign}
We know that in $\Lambda$CDM matter-dominated universe $\Omega_{\rm m}\sim 1$, $H=2/(3t)$ and $f_{\rm m}^{\prime}=0$, so $f_{\rm m}=1$.
Hence we can find an approximate solution for the deviation of the growth rate from $\Lambda$CDM,
\begin{equation}
\begin{aligned}
    \delta f \simeq &\frac{5}{2}\,e^{-5/2\ln a}\int{e^{5/2\ln a}\delta\mu_{\Psi}\,d\ln a}\\
    &\propto \delta\mu_\Psi\,,
    \label{equ:deltaf}
\end{aligned}
\end{equation}
where to obtain the second line we have assumed that $\delta\mu_\Psi \sim \text{const.}$, which is reasonable in matter domination where in most realistic models the $\alpha_i$ are negligible.

We plot this result in Figure~\ref{fig:growth} and compare it to the one obtained with \texttt{hi\_class}.
We can see that in the matter-dominated era, the two solutions agree very well and the approximation fails at later times, where the assumptions we have made do not hold.

\subsection{CMB power spectra}\label{sec:cmb}
\begin{figure}
    \centering
    \begin{subfigure}[t]{0.48\textwidth}
        \includegraphics[width=\textwidth]{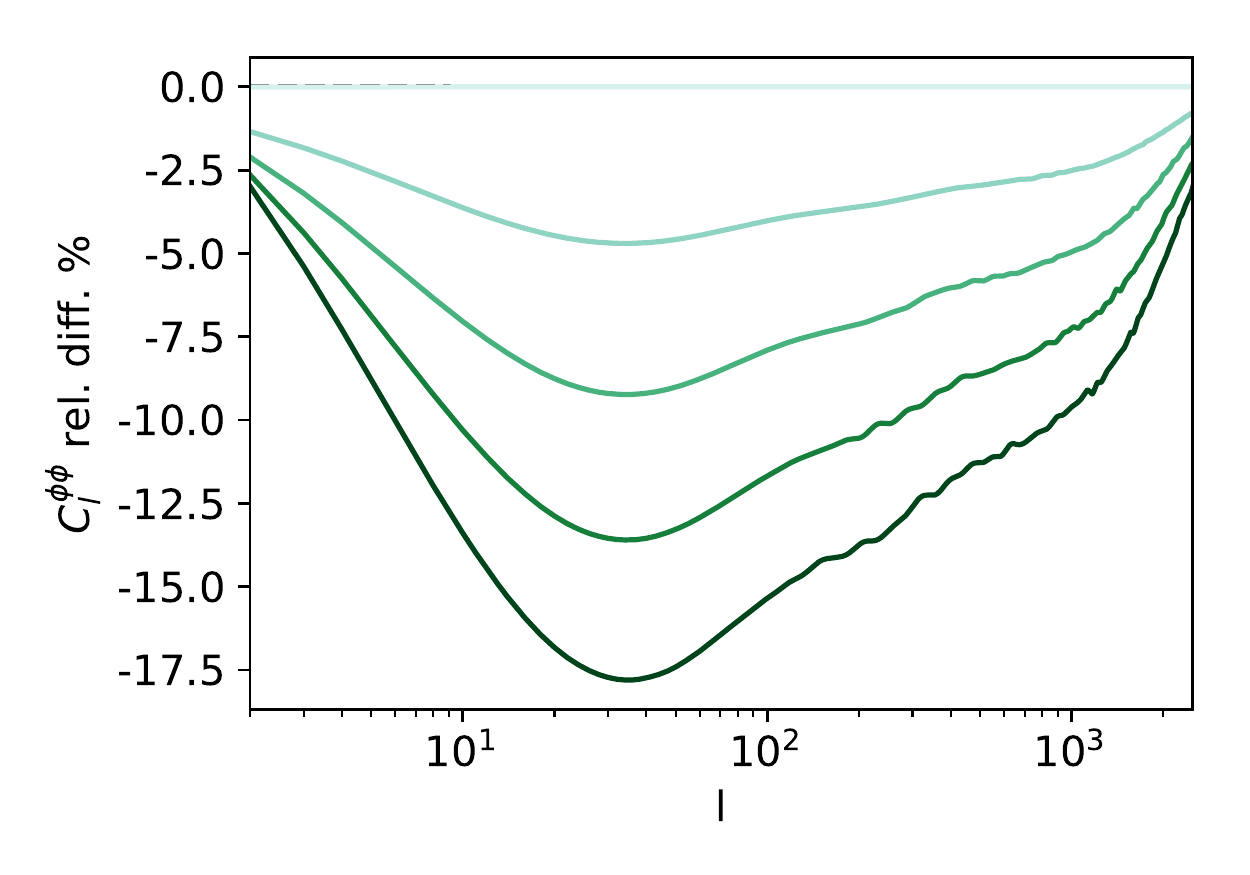}
    \end{subfigure}
    \,
    \begin{subfigure}[t]{0.48\textwidth}
        \includegraphics[width=\textwidth]{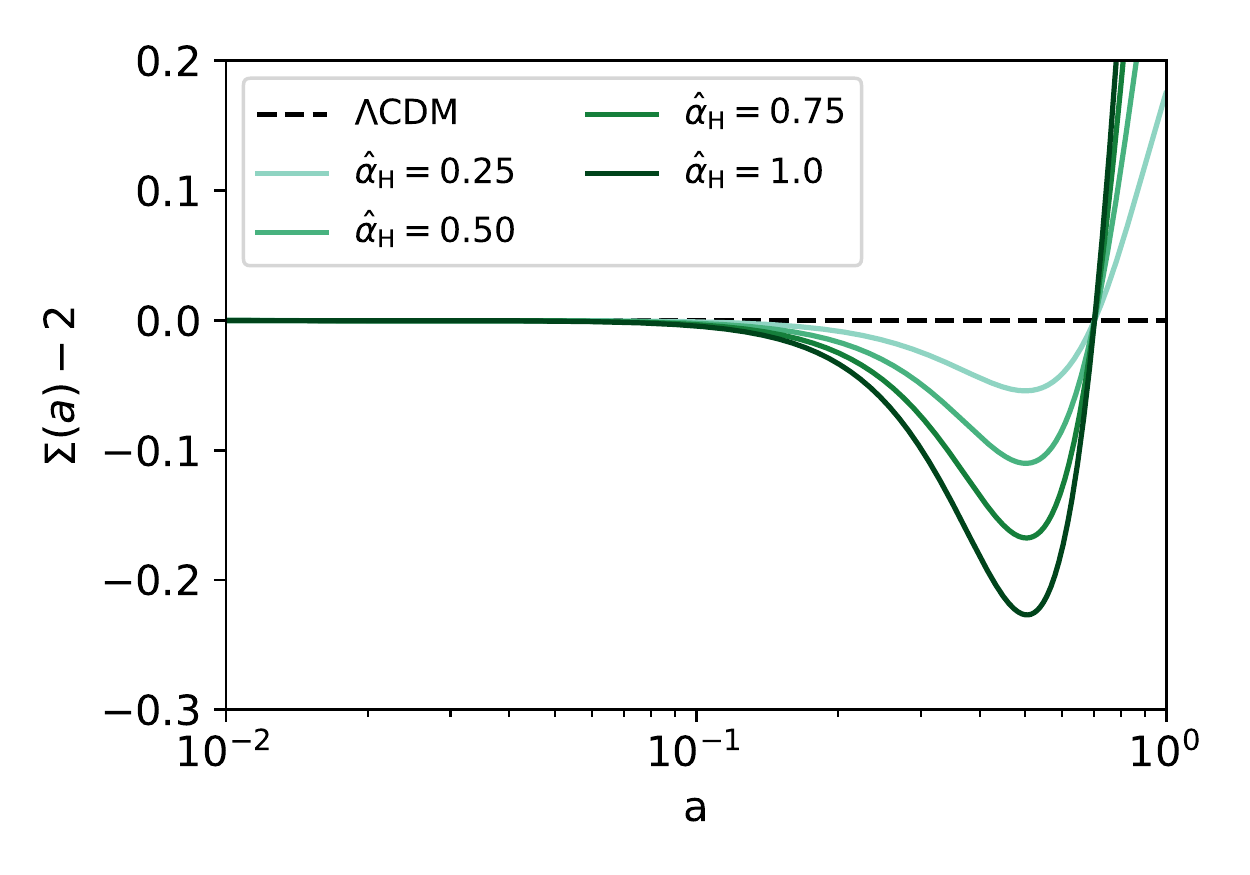}
    \end{subfigure}
    \caption{On the left is the relative difference of the CMB lensing potential for different values of $\hbeh$ and $\Lambda$CDM. And on the right is the deviation from $\Lambda$CDM of the lensing parameter, $\Sigma(a) - 2$. The remaining parameters of the model are set as above.}
    \label{fig:lcl}
\end{figure}
\begin{figure}
    \centering
    \begin{subfigure}[t]{0.45\textwidth}
        \includegraphics[width=\textwidth]{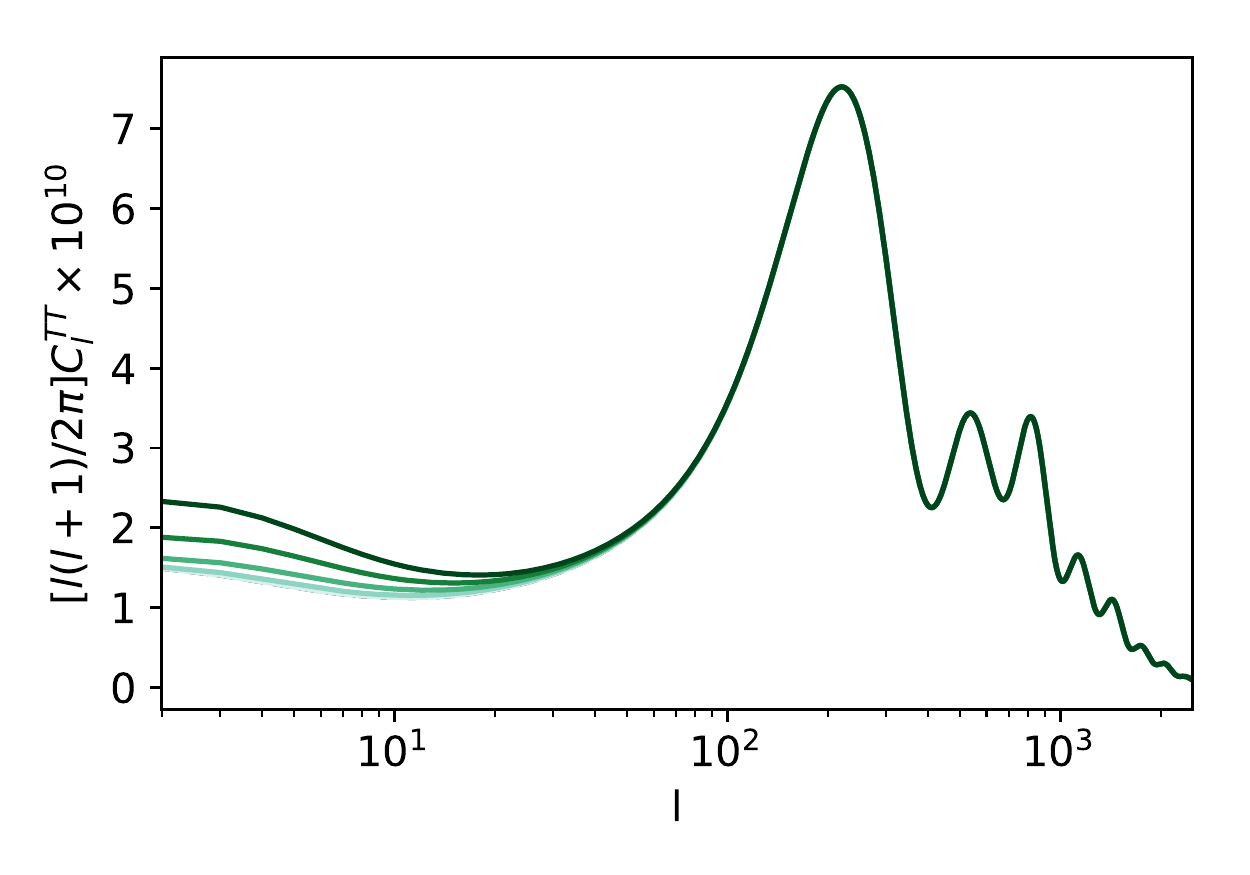}
    \end{subfigure}
    \,
    \begin{subfigure}[t]{0.45\textwidth}
        \includegraphics[width=\textwidth]{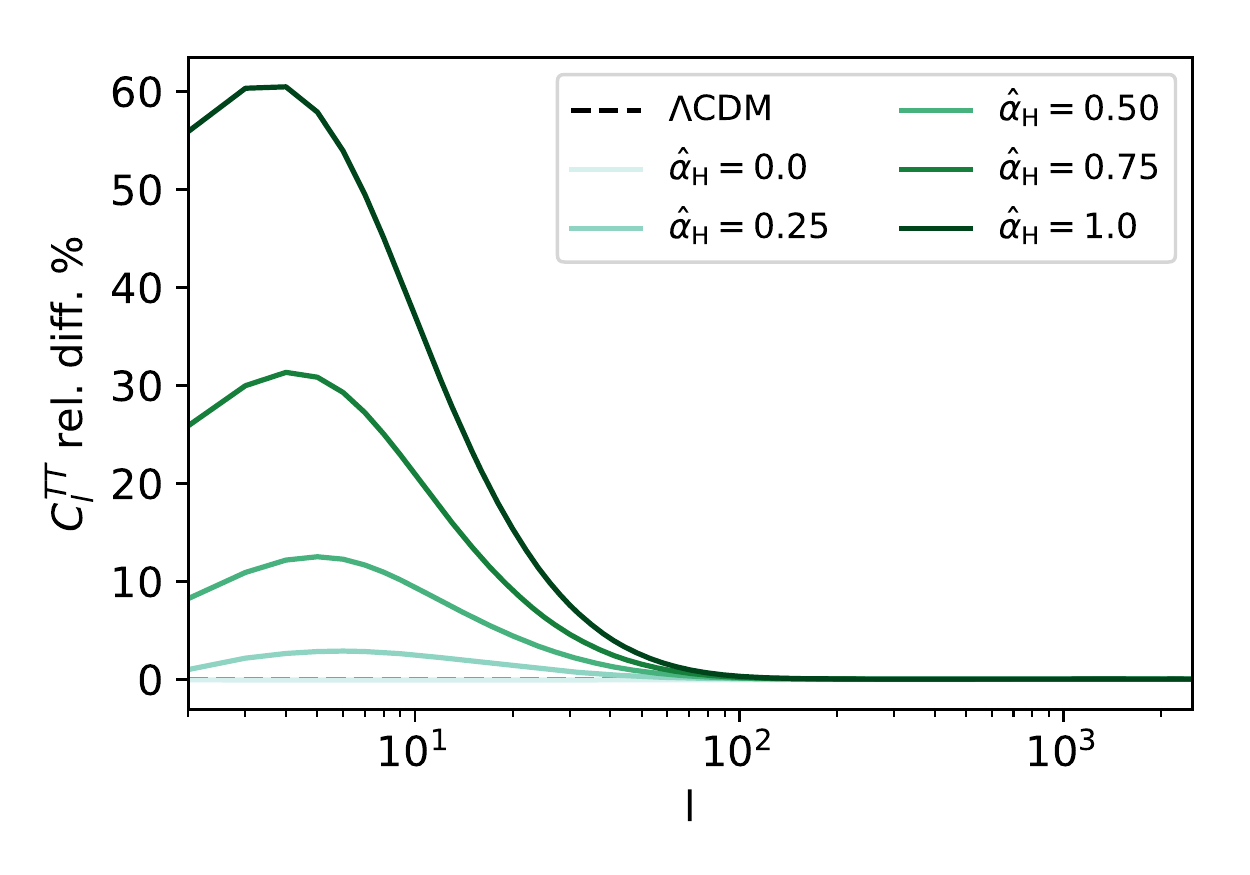}
    \end{subfigure}
        \\
    \centering
    \begin{subfigure}[t]{0.45\textwidth}
        \includegraphics[width=\textwidth]{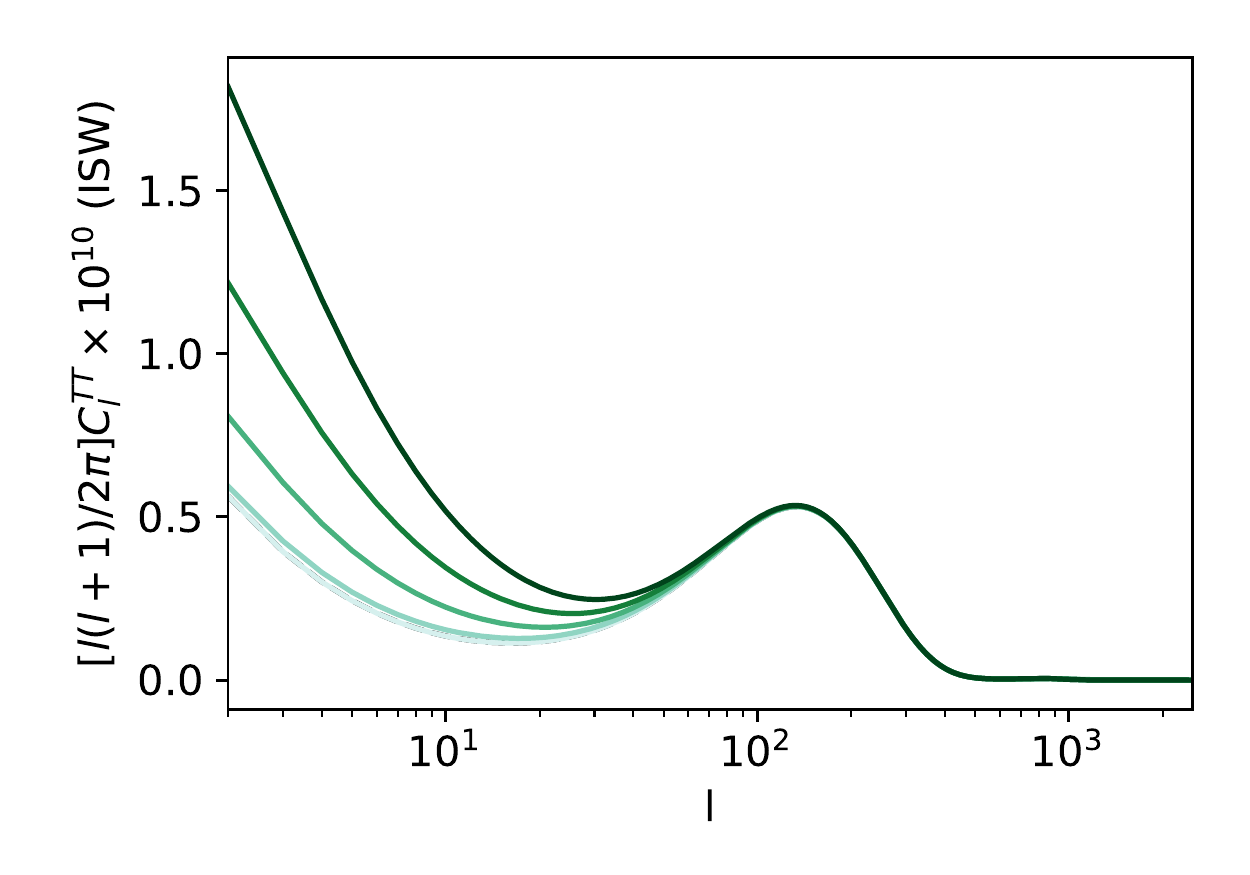}
    \end{subfigure}
    \begin{subfigure}[t]{0.45\textwidth}
        \includegraphics[width=\textwidth]{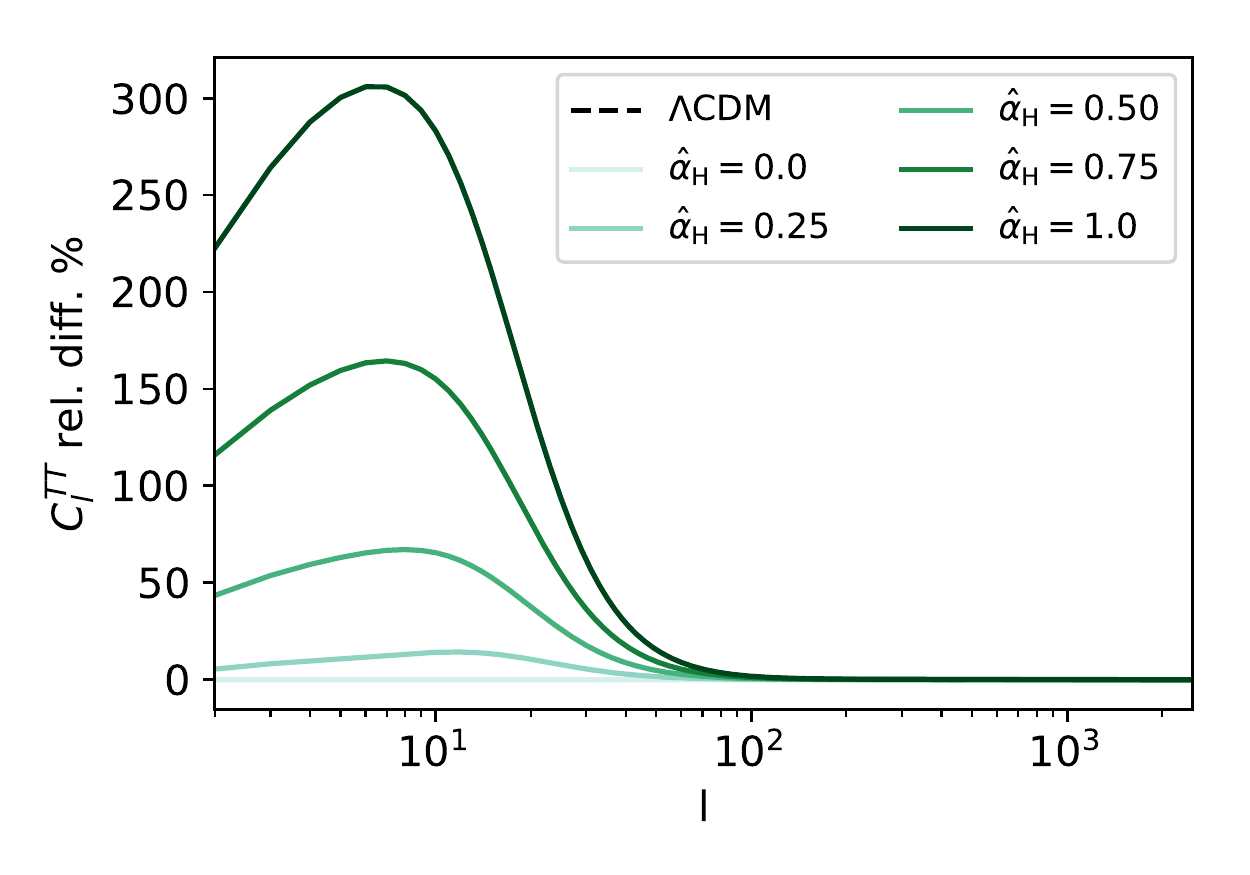}
    \end{subfigure}
    \caption{The unlensed CMB temperature power spectrum for different values of $\beh$, compared to $\Lambda$CDM.}
    \label{fig:tcl}
\end{figure}
The fluctuations in the CMB temperature and polarisation can be used to constrain very tightly the standard cosmological parameters in the $\Lambda$CDM model.
Large scale structure gravitationally lenses the temperature and polarisation anisotropies, which makes the CMB also a good probe of the late universe where the effects that we are interested in become important. 
Therefore, beyond Horndeski theories are expected to modify the CMB spectra. In particular we expect to see significant deviations from $\Lambda$CDM in the lensing potential and the temperature power spectrum of the CMB.
On the left panel of Figure~\ref{fig:lcl} we show the relative difference between the CMB lensing potential, $C_l^{\phi\phi}$, of a set of MG models and our fiducial $\Lambda$CDM.
As before we are looking at models where all $\hbra=\hrun=\hten=0$ and $\hkin=1$ and consider five different values of $\hbeh$.
We can see that $C_l^{\phi\phi}$ decreases at a nearly linear rate as we increase $\hbeh$, which was also shown in \cite{DAmico:2016ntq}.
Again, in order to understand this behaviour, we turn to the equations describing this effect in the quasi-static approximation.
The strength of the lensing potential is related to the Weyl potential, which can be obtained by summing the constraint Eqs.~(\ref{equ:QSphi}) and (\ref{equ:QSpsi}),
\begin{equation}
    \frac{k^2}{a^2}(\Phi+\Psi)=-\frac{3}{2}H^2\Omega_{\rm m}(\mu_\Phi+\mu_\Psi)\delta_{\rm m} + \beh H (\lambda_\Phi + \lambda_\Psi)\dot{\delta}_{\rm m}\,.
\end{equation}
Similarly to the case of the power spectrum, Section~\ref{sec:mpk}, we look at the simplest case (where $\bra=\run=\ten=0$). Here we have $\mu_\Phi$ and $\mu_\Psi$ given by Eqs.~(\ref{equ:la-mu_phi}) and (\ref{equ:la-mu_psi}) respectively.
As in the case with the effective gravitational constant, the weak lensing parameter can not be defined as a function of $\mu_\Psi$ and $\mu_\Phi$ as usual.
Here, instead we have
\begin{equation}
\begin{aligned}
    \Sigma\equiv -\frac{2k^2(\Phi+\Psi)}{3a^2H^2\Omega_{\rm m}\delta_{\rm m}}
        &=\mu_\Phi + \mu_\Psi - \frac{2\beh(\lambda_\Phi + \lambda_\Psi)}{3\Omega_{\rm m}}f\,,
\end{aligned}
\end{equation}
where again we have used $\dot{\delta} = fH\delta$.
For the case of $\hbra=\hkin=\hten=0$ and $\hrun=1$ the above expression reduces to
\begin{equation}
    \Sigma = 2 - \beh(\lambda_\Phi + \lambda_\Psi)\left(1+\frac{2f}{3\Omega_{\rm m}}\right)\,.
\end{equation}
This expression is plotted on the right panel of Figure~\ref{fig:lcl}.
In $\Lambda$CDM the lensing potential is $\Sigma = 2$, so what we are showing is the modification to this quantity due to the beyond Horndeski parameter, $\beh$.
It is possible to notice that this is negative during most of the matter dominated era and decreases as $\beh$ increases, which confirms the damping effect of $\beh$ on the lensing potential, see left panel of Figure~\ref{fig:lcl}.

The effects of beyond Horndeski can be seen also on the top panel of Figure~\ref{fig:tcl}, where we plot the CMB temperature power spectrum as obtained by \texttt{hi\_class}.
This result is again consistent with \cite{DAmico:2016ntq}.
We can see that increasing $\hbeh$ enhances the CMB temperature power spectrum at low $l$.
This effect is well known for DE/MG models, and it is due to the integrated Sachs-Wolfe effect, which affects the power spectrum at low $l$, but has no significant effect at large $l$. The contribution from the combination of the late and early integrated Sachs-Wolfe effect to the temperature anisotropies is plotted on the bottom of Figure~\ref{fig:tcl}.

\section{Results}\label{sec:results}
In this section we present the constraints on our specific choice of beyond Horndeski parameters with data. We note that different time dependences may lead to large changes in the uncertainties and that a more systematic analysis including a broader range (or variety) of priors would be desirable. Nevertheless, this can be seen as a first step in assessing constraints on par with what has been done in the case of Horndeski theories.
\subsection{Method}
We use the Boltzmann code \texttt{hi\_class} \cite{Zumalacarregui:2016pph} to solve the equations shown in Appendix \ref{app:hi-class}.
This code extends the Cosmic Linear Anisotropy Solver Software (CLASS) \cite{2011JCAP...07..034B} by including the Horndeski class of theories, and it has been tested intensively against other Einstein-Boltzmann solvers in \cite{Bellini:2017avd}.
On top of the public version of \texttt{hi\_class}, we have added the extra beyond Horndeski terms to the perturbation equations, the definitions of the DE density, pressure and $\beh$.
We then interfaced it with the modular cosmological parameter estimation code CosmoSIS \cite{Zuntz:2014csq}, which runs Monte Carlo Markov Chains (MCMC).
We used the Metropolis-Hastings sampler to obtain our chains and consider those converged when the Gelman and Rubin parameter $R-1<0.01$  \cite{Gelman:1992zz}.

The full set of beyond Horndeski parameters that describes the evolution of the perturbations is $\{\hkin,\hbra,\hrun,\hten,\hbeh\}$.
Assuming the constraints derived from the GW detection that we discussed in Section~\ref{sec:gw17}, we set $\hten=0$ for all runs and run two sets of MCMC chains, one fixing $\hkin=1$ and another with $\kin$ given by Eq.~(\ref{equ:kin_fix}) to ensure $\cs=1$ for all possible sets of values of $\{\hbra,\hrun,\hbeh\}$. 
We fix the initial effective Planck mass, $\Ms=1$.
We also assume two massless and one massive neutrino with $m_\nu=0.06$eV and vary the background cosmological parameters and the relevant set of $\hat{\alpha}_i$.
Given that we use a parametrisation for the evolution of the $\alpha_i$, Eq.~(\ref{equ:propto_omega}), here we present the constraints for the coefficients that fix this proportionality, $\hat{\alpha}_i$.

\subsection{Datasets}
\begin{table}
\begin{center}
\begin{tabular}{ |p{3cm}||p{3cm}|p{3cm}|  }
 \hline
 \multicolumn{3}{|c|}{BAO measurements} \\
 \hline
 Survey     & $z$   &  $D_{\rm V}$ \\
 \hline
 6dFGS      \cite{2011MNRAS.416.3017B} & $0.106$ & $456\pm27$\\
 SDSS-MGS   \cite{Ross:2014qpa}        & $0.15$  & $664\pm25$\\
 BOSS DR12  \cite{Alam:2016hwk}     & $0.38$  & $1477\pm16$\\
 BOSS DR12 \cite{Alam:2016hwk}    & $0.51$  & $1877\pm19$\\
 BOSS DR12 \cite{Alam:2016hwk}    & $0.61$  & $2140\pm22$\\
 \hline
\end{tabular}
\caption{List of BAO measurements used in this work.}
\label{tab:bao}
\end{center}
\end{table}
\begin{table}
\begin{center}
\begin{tabular}{ |p{3cm}||p{3cm}|p{3cm}| }
 \hline
 \multicolumn{3}{|c|}{RSD measurements} \\
 \hline
 Survey     & $z$   & $f(z)\sigma_8(z)$\\
 \hline
 6dFGS   \cite{2012MNRAS.423.3430B}  & $0.067$ & $0.423\pm0.055$\\
 SDSS-MGS     \cite{Howlett:2014opa}      & $0.15$  & $0.53\pm0.19$\\
 SDSS-LRG     \cite{Oka:2013cba}          & $0.30$  & $0.49\pm0.09$\\
 WIGGLEZ \cite{2011MNRAS.415.2876B}  & $0.22$  & $0.42\pm0.07$\\
 WIGGLEZ \cite{2011MNRAS.415.2876B}  & $0.41$  & $0.45\pm0.04$\\
 WIGGLEZ \cite{2011MNRAS.415.2876B}  & $0.60$ & $0.43\pm0.04$\\
 WIGGLEZ \cite{2011MNRAS.415.2876B}  & $0.78$  & $0.38\pm0.04$\\
 BOSS DR12     \cite{Alam:2016hwk}          & $0.38$  & $0.497\pm0.032$\\
 BOSS DR12     \cite{Alam:2016hwk}          & $0.51$  & $0.458\pm0.025$\\
 BOSS DR12     \cite{Alam:2016hwk}          & $0.61$  & $0.436\pm0.022$\\
 VIPERS  \cite{Pezzotta:2016gbo}    & $0.60$ & $0.55\pm0.12$\\
 VIPERS  \cite{Pezzotta:2016gbo}    & $0.86$  & $0.40\pm0.11$\\
 \hline
\end{tabular}
\caption{List of RSD measurements used in this work.}
\label{tab:rsd}
\end{center}
\end{table}

\begin{figure}
    \centering
    \begin{subfigure}[t]{0.48\textwidth}
        \includegraphics[width=\textwidth]{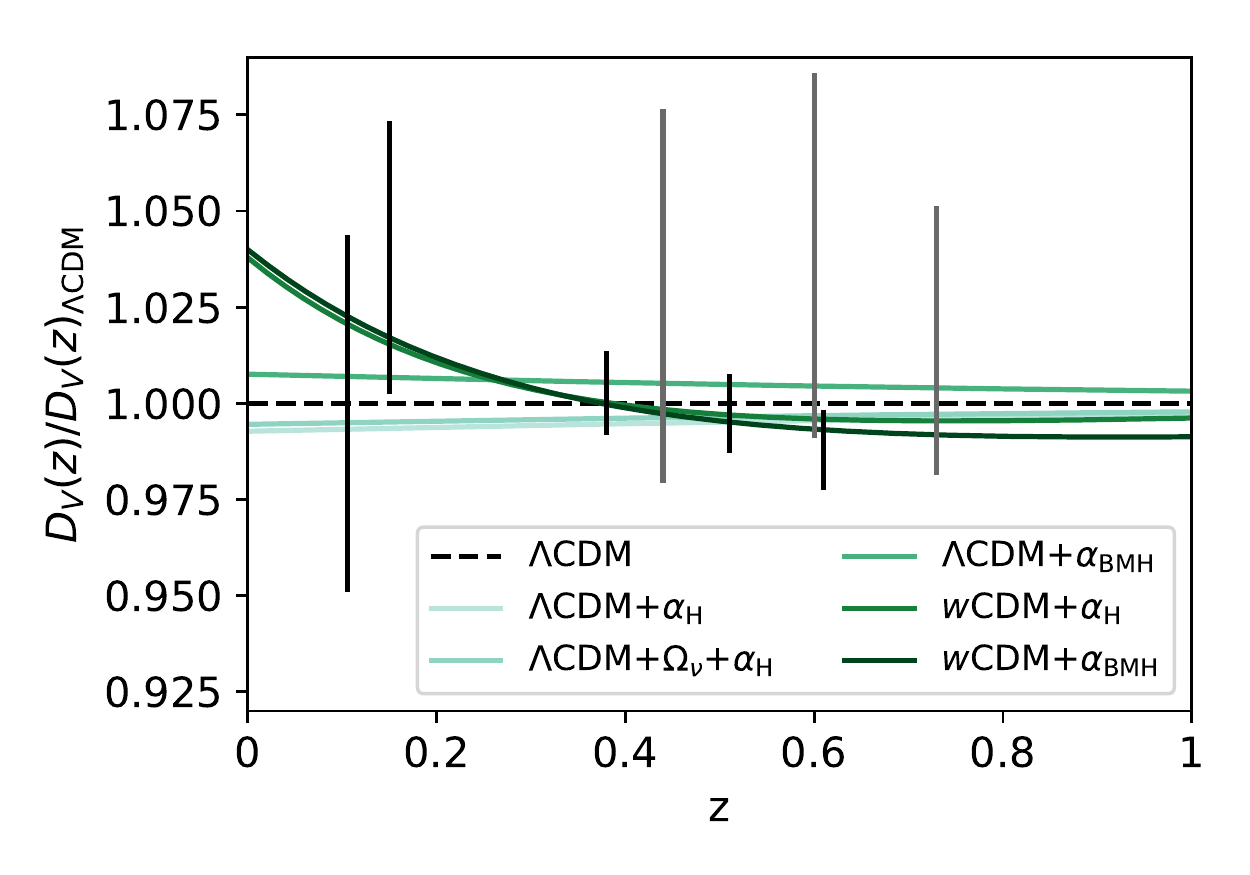}
    \end{subfigure}
    \,
    \begin{subfigure}[t]{0.48\textwidth}
        \includegraphics[width=\textwidth]{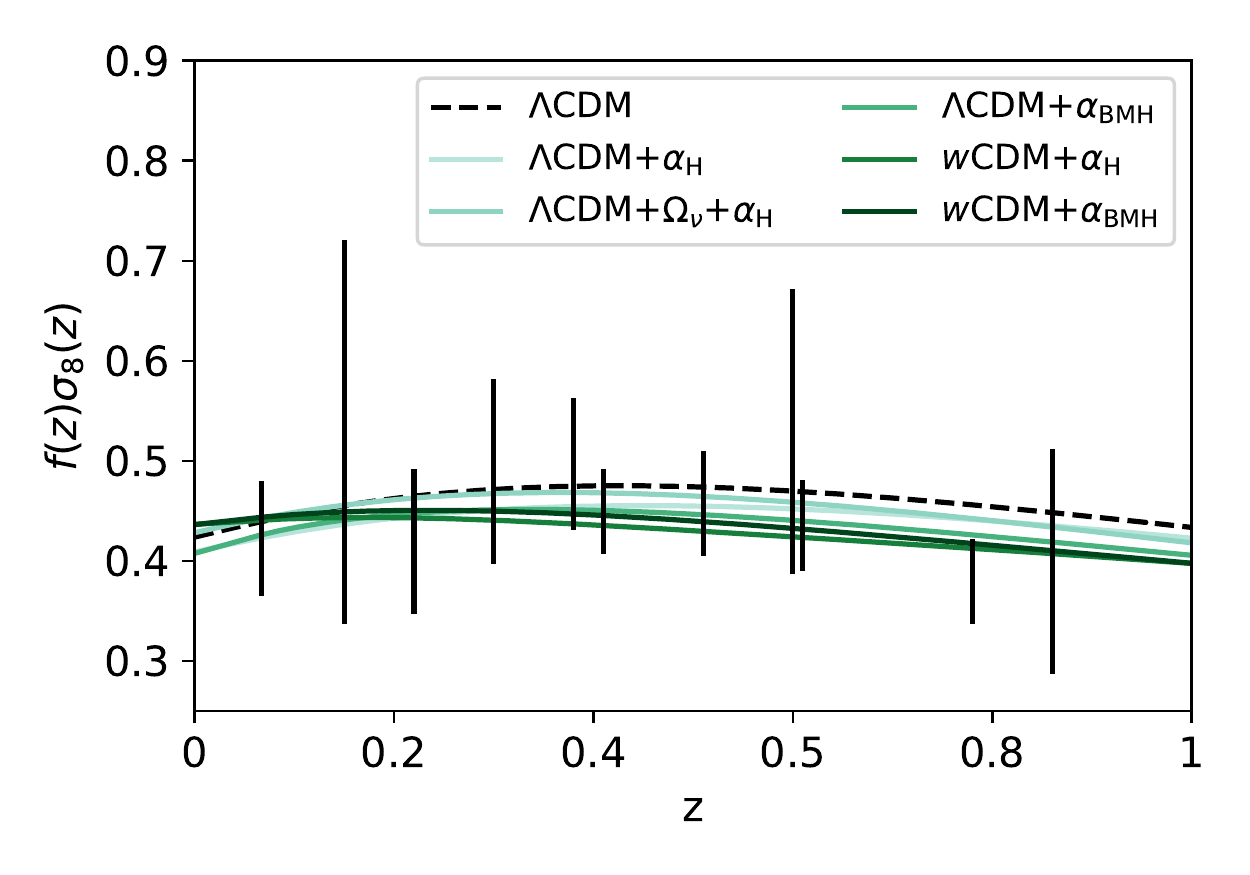}
    \end{subfigure}
    \caption{Distance measurements compared to $\Lambda$CDM (left) and the growth of structure $f\sigma_8$ (right) as a function of redshift, z, for best fit lines of a set of MG models. 
    On the left plot, the black lines are the BAO measurements from Table~\ref{tab:bao}and the grey are the measurements from the WiggleZ survey \cite{2011MNRAS.415.2876B}. 
    On the right are the RSD measurement from Table~\ref{tab:rsd}.}
    \label{fig:bao-rsd}
\end{figure}

To constrain the parameters of the theory we include CMB data, measurements of baryon acoustic oscillations (BAO) and redshift space distortions (RSD) from large-scale structure surveys:
\begin{itemize}
    \item CMB: We use the temperature and polarisation power spectra and the lensing potential of the Cosmic Microwave Background (CMB) from Planck 2015 \cite{Adam:2015rua,Ade:2015xua}. 
    The Planck 2015 likelihood is discussed in \cite{Aghanim:2015xee}. We use the high $l$ TT likelihood with $l=30-2508$, along with the joint TT, EE, BB and TE likelihood in the range $l=2-29$ and the lensing likelihood, using both temperature and polarisation lensing reconstruction in the multipole range $l=40-400$.

    \item BAO: 
    We use BAO measurements from 6dFGS \cite{2011MNRAS.416.3017B} and the Sloan Digital Sky Survey (SDSS) Data Release 7 Main Galaxy Sample (MGS) \cite{Ross:2014qpa}, which probes the supplied expansion history through the redshift-distance and redshift-Hubble relations combined through the relation, $D_{\rm V} = (D_{\rm A}^2(1+z)^2z/H)^{1/3}$. 
Here $D_{\rm V}$ is the angle-averaged distance, $D_{\rm A}$ - the angular diameter distance and $z$ is the redshift.
We also used measurements from the Baryon Oscillation Spectroscopic Survey (BOSS) Data Release 12 \cite{Alam:2016hwk}, which are lower signal-to-noise and constrain both the angular diameter distance, $D_{\rm A}$ and the Hubble parameter, $H$.
There are also BAO measurements from WiggleZ survey \cite{2011MNRAS.415.2876B}, but these partially overlap with the BOSS CMASS volume. We have chosen to use the BOSS measurements here.
These measurements do not constrain the matter power spectrum itself, but just the positions of the BAO peaks.
They are presented in Table~\ref{tab:bao} and plotted on the left panel of Figure~\ref{fig:bao-rsd}.

    \item RSD: Finally, we use Redshift Space Distortions (RSD) data. This effect probes the growth of structure and therefore is sensitive to late-time effects and can be used to probe MG.
We use measurements derived form from 6dFGS \cite{2012MNRAS.423.3430B}, SDSS MGS \cite{Howlett:2014opa} and Luminous Red Galaxy (LRG) \cite{Oka:2013cba} samples, WiggleZ \cite{2011MNRAS.415.2876B}, and VIPERS \cite{Pezzotta:2016gbo}.
We also use the RSD measurements from BOSS DR12 \cite{Alam:2016hwk} with the full covariance between the 3 $f(z)\sigma_8(z)$ measurements at different redshifts and the BAO measurements of $H(z)$ and $D_{\rm A}(z)$.
All RSD measurements are quoted on Table~\ref{tab:rsd} and on the right panel of Figure~\ref{fig:bao-rsd}.
\end{itemize}
In \cite{Bellini:2015xja} the authors also included data for the full matter power spectrum, $P(k)$, from a number of large scale structure surveys.
However, this dataset does not improve the constraints significantly and requires to choose a galaxy bias factor. So we have decided not to include it here.
\subsection{Constraints on the beyond Horndeski parameters}\label{sec:constr}
\begin{table}
\begin{center}
\begin{tabular}{ |p{3cm}|p{3.5cm}|p{3.7cm}|p{3.5cm}|}
\hline
 & Data & Constraints & Marg. mean\\
 \hline
 $\Lambda$CDM+$\beh$ & CMB+BAO  &  $0.01<\hbeh<0.31$  &  $\hbeh = 0.12 \pm 0.10$\\
   & CMB+BAO+RSD &  $0.02<\hbeh<0.36$  &  $\hbeh = 0.17\pm 0.11$\\
 \hline
 $\Lambda$CDM+$\Omega_\nu$+$\beh$ & CMB+BAO & $0.01<\hbeh<0.33$ & $\hbeh = 0.12 \pm 0.10$ \\ 
  & CMB+BAO+RSD & $0.01<\hbeh<0.33$ &  $\hbeh = 0.14 \pm 0.10$ \\ 
 \hline
 $\Lambda$CDM+$\alpha_{\rm BMH}$ & CMB+BAO  & $-0.44<\hbeh<2.16$ &  $\hbeh = 0.96 \pm 0.80$\\
   & CMB+BAO+RSD & $0.38<\hbeh<2.48$  &  $\hbeh = 1.49\pm 0.64$\\
 \hline
 $w$CDM+$\beh$  & CMB+BAO & $0.13<\hbeh<0.51$ &  $\hbeh = 0.35 \pm 0.11$\\
   & CMB+BAO+RSD & $0.02<\hbeh<0.49$ &  $\hbeh = 0.33 \pm 0.12$\\
 \hline
 $w$CDM+$\alpha_{\rm BMH}$  & CMB+BAO & $-0.53<\hbeh<2.18$ &  $\hbeh = 0.90 \pm 0.69$ \\
    & CMB+BAO+RSD & $0.07<\hbeh<2.21$ &  $\hbeh = 1.15 \pm 0.65$\\
 \hline
\end{tabular}
\caption{Constraints (95\%) and marginalised mean values of $\beh$ for the different datasets combinations and different sets of other parameters being varied.}
\label{tab:grid}
\end{center}
\end{table}
Here we present the constraints on the beyond Horndeski parameters derived using the method and datasets described above.
As mentioned before, the value of $\hkin$ has little to no effect on sub-horizon scales and in particular on the observables we are looking at.
We then chose either to fix $\hkin=1$ or to get it from Eq.~(\ref{equ:kin_fix}) and present the constraints on the other parameters.

The grid of parameters varied and datasets used is shown on Table~\ref{tab:grid}, with the 95\% confidence regions for $\hbeh$ and marginalised means.
Our baseline model is the standard $\Lambda$CDM, where we varied the Hubble constant $H_0$, the baryon and dark matter densities today, $\Omega_{\rm b}$ and $\Omega_{\rm cdm}$, the curvature fluctuation amplitude, $A_{\rm s}$, the scalar spectral index, $n_{\rm s}$ and the optical depth at reionization, $\tau$.
On top of that, in the other runs we varied: the beyond Horndeski parameter $\hbeh$ keeping the other $\hat{\alpha}_i$ fixed ($\Lambda$CDM+$\beh$); $\hbeh$ and the density of massive neutrinos ($\Lambda$CDM+$\Omega_\nu$+$\beh$); $\hbeh$ and the Horndeski $\alpha_i$ ($\Lambda$CDM+$\alpha_{BMH}$); $\hbeh$ and we replaced the cosmological constant $\Lambda$ with a time dependent equation of state parametrized as $w(a)= w_0 + w_{a}(1 - a)$ ($w$CDM+$\alpha_{H}$); and finally a time dependent equation of state plus the Horndeski $\hat{\alpha}_i$ and $\hbeh$ ($w$CDM+$\alpha_{BMH}$).

\begin{figure}
    \centering
        \includegraphics[width=0.5\textwidth]{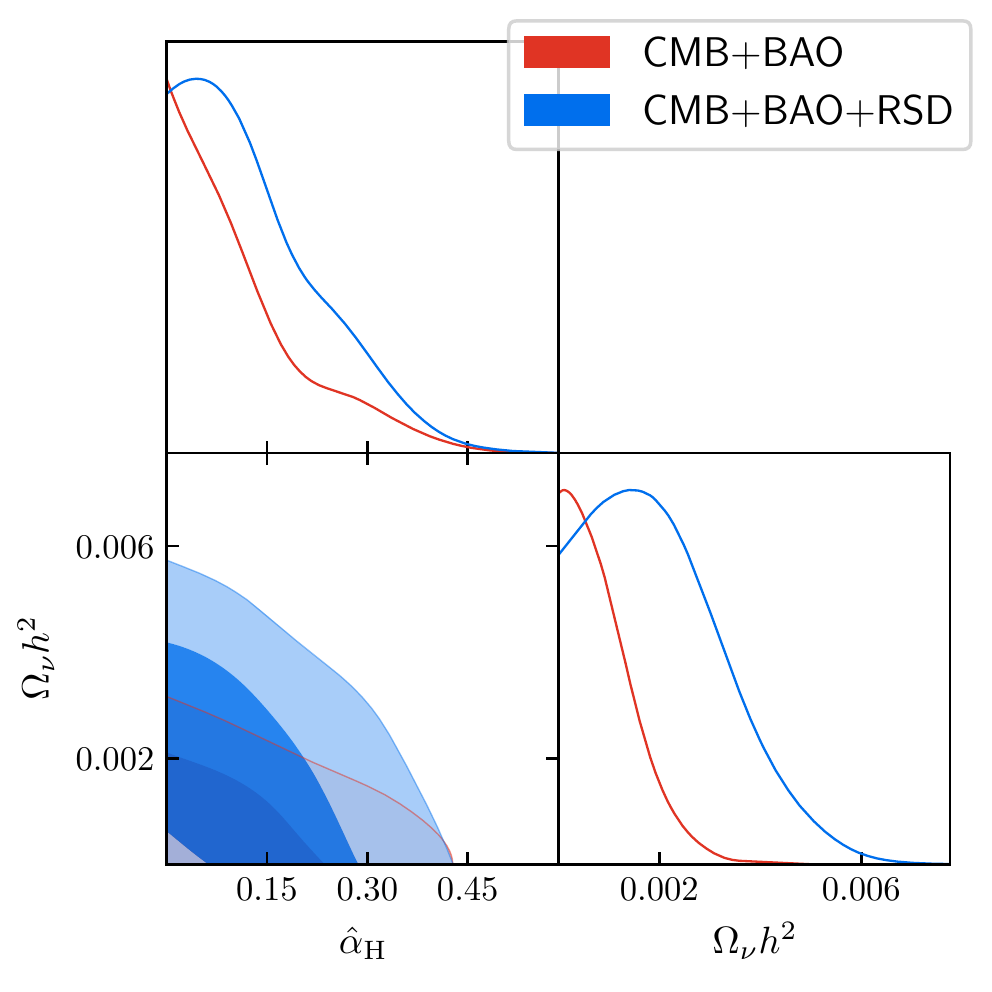}
     \caption{Constraints on $\Omega_\nu$ and $\hbeh$ from the combination of CMB+BAO (red) and CMB+BAO+RSD (blue) datasets from the $\Lambda$CDM+$\Omega_\nu$+$\beh$ run. We evolve the $\beh$ as proposed \cite{Bellini:2014fua}, with $\hkin =1$, $\hbra=\hrun=\hten=0$.}
     \label{fig:hnu}
\end{figure}

\begin{figure}
    \centering
        \includegraphics[width=\textwidth]{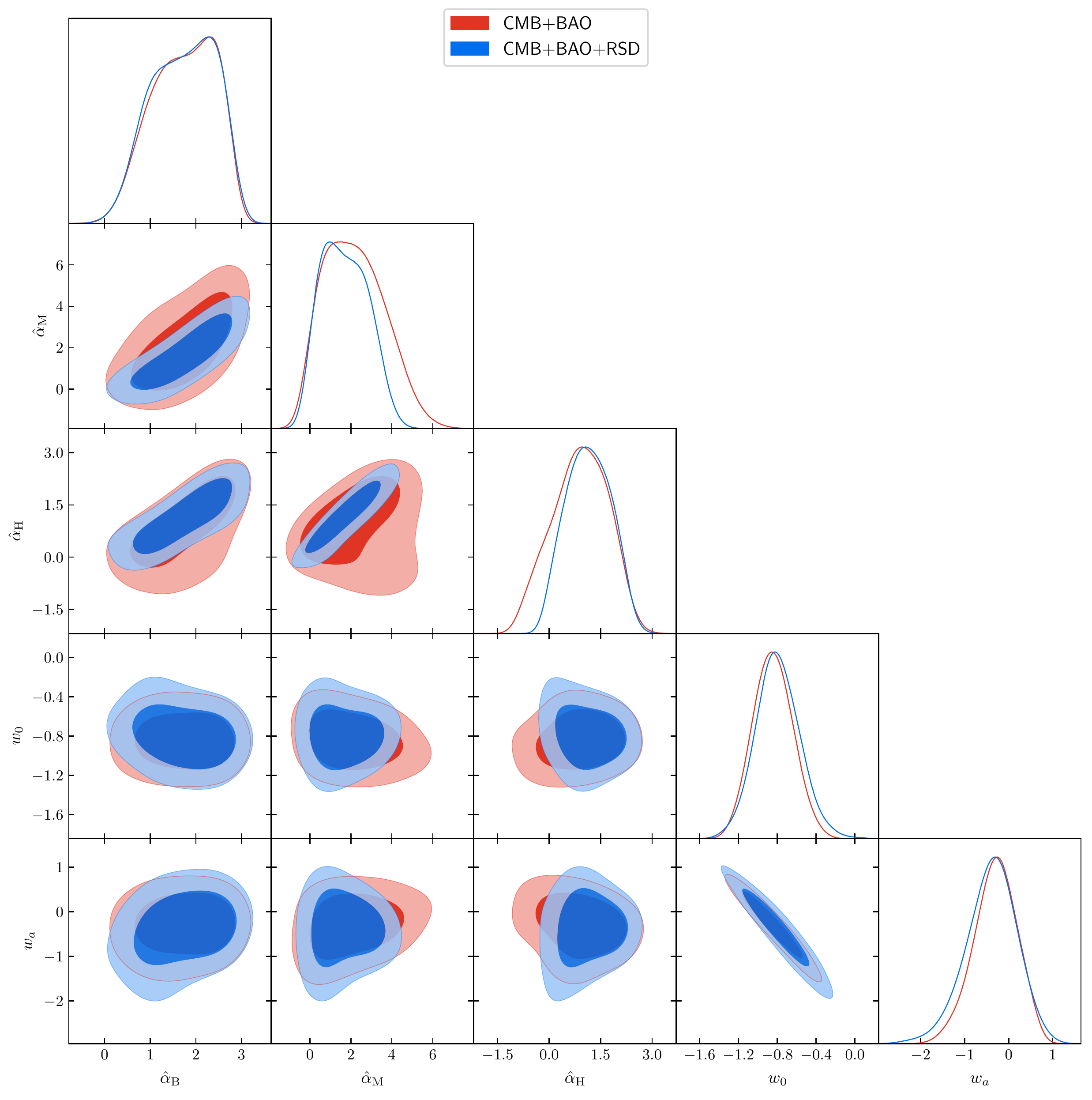}
     \caption{Constraints on the MG parameters, ${\hbra,\hrun,\hbeh}$, and the DE equation of state parameters, $w_0$ and $w_a$. These contours are from the full parameter run ($w$CDM+$\alpha_{\rm BMH}$) with the combination of CMB+BAO (red) and CMB+BAO+RSD (blue) datasets.}
     \label{fig:bmhw}
\end{figure}

The confidence regions of $\hbeh$ seem to exclude $0$ in most of these cases and are mainly positive.
The lower limit at $0$ is exact when considering a cosmological constant and $\hbra=\hrun=0$, and can be inferred by inspecting Eq.~(\ref{equ:speed}).
For the remaining cases the contours are smooth around $0$ since the other MG parameters modify the stability conditions, but Eq.~(\ref{equ:speed}) still plays a central role in excluding negative values of $\hbeh$. 
Both the upper and lower limits seem to be affected largely by the combination of datasets used and the freedom in some of the other parameters.
We see that varying $\Omega_\nu$ has little effect on the constraints of $\hbeh$ and the largest difference seems to be due to the other MG parameters $\hbra$ and $\hrun$.
It is possible to notice from Table~\ref{tab:grid}, and Figures~\ref{fig:hnu} and \ref{fig:bmhw}, that the contours of the MG parameters are greatly improved by adding the RSD dataset.
This can be explained by the fact that RSD is the only dataset that probes the growth rate of structure, $f\sigma_8$, at high significance.
And given that we chose for the $\alpha_i$ to evolve as the dark energy density parameter, $\Omega_{\rm DE}$, we would expect it to have an effect at late times.

\begin{table}
\begin{center}
\begin{tabular}{ |p{3.2cm}|p{3cm}|p{3cm}|p{3cm}| }
\hline
  Model & CMB  & BAO+RSD & Total \\
 \hline
 $\Lambda$CDM  & 5636.61 (--)      & 4.78 (--)    & 5641.39 (--)\\
 $\Lambda$CDM+$\beh$  & 5636.05 (0.56) & 3.55 (1.23) &  5639.60 (1.79)\\
 $\Lambda$CDM+$\Omega_\nu$+$\beh$  &  5636.29 (0.35) & 3.72 (1.03) & 5640.01 (1.38)\\
 $\Lambda$CDM+$\alpha_{BMH}$  & 5633.84 (2.77) & 2.95 (1.83) & 5636.79 (4.60)\\
 $w$CDM+$\alpha_{H}$  & 5634.19 (2.42) & 3.27 (1.51) & 5637.46 (3.93)\\
 $w$CDM+$\alpha_{BMH}$  &  5632.62 (3.99) & 2.82 (1.96) &  5635.44 (5.95) \\
 \hline
\end{tabular}
\caption{Best fit likelihoods for each model and probe, when running the full CMB+ BAO+RSD dataset. In the brackets we show the difference of the likelihood for the given model from $\Lambda$CDM.}
\label{tab:like}
\end{center}
\end{table}
In Table~\ref{tab:like} we show the absolute log likelihood values for the models considered, measured at their best fit values.
It is possible to notice that the total likelihood generally decreases as we increase the number of free parameters.
A better fit to data is expected when enlarging the parameter space, but in this case the improvement does not appear enough to justify the price of adding new parameters.
The best fit models are also plotted, together with the measurements we used, on  the left panel of Figure~\ref{fig:bao-rsd} for the BAO datasets and on the right for the RSD dataset.
We show how these measurements are fitted by the models we have considered here (green lines) and $\Lambda$CDM (black dashed lines). 
The beyond Horndeski models, and particularly when combined with the DE equation of state clearly provide more flexibility to fit these measurements than the best likelihood model of $\Lambda$CDM.
However, given that most of these models fit within the error bars of these measurements, we cannot conclude that these differences point towards either of the models being a better fit.

On Figure~\ref{fig:hnu} we have plotted the confidence regions for the relevant parameters of the  $\Lambda$CDM+$\Omega_\nu$+$\beh$ run.
The lighter and darker contours represent the 95\% and 68\% confidence regions respectively.
In this case the only MG parameter we vary is $\hbeh$, while the remaining $\hat{\alpha}_i$ are fixed to $\hkin = 1$, $\hbra=\hrun=\hten=0$. As mentioned the gradient stability condition, Eq.~(\ref{equ:speed}), requires $\hbeh>0$, which can be clearly seen in the contours. Interestingly, $\hbeh$ does not seem to have any degeneracy with the density of neutrinos $\Omega_\nu$, which was the main motivation to study this extension to the baseline $\Lambda$CDM model.

Figure~\ref{fig:bmhw} shows the contours for the $w$CDM+$\alpha_{BMH}$ model. In particular, we focus on the posterior distributions of the (beyond) Horndeski parameters,  $\{\hbra,\hrun,\hbeh\}$, and the DE EoS parameters, $w_0$ and $w_a$. As mentioned, by comparing the two dataset combinations, CMB+BAO and CMB+BAO+RSD, it is possible to notice the effect of RSD in tightening the contours.
The most significant differences seem to be for $\hkin$ and $\hbeh$.
Unlike in the previous case, here there is no hard bound on $0$ on the value of $\hbeh$ coming from the stability conditions. This allows it to go negative, since the combination with the other parameters may preserve the stability of the theory.
In the case where we had $\hbra=\hrun = 0$, the data seem to peak the distribution of $\hbeh$ around zero and no large values were allowed. On the contrary here, the degeneracy between $\hbeh$, $\hbra$ and $\hrun$, spreads out the distribution of $\hbeh$ in the positive direction relaxing significantly the contours. We see that $\beh$ is not degenerate with the $w_{\rm DE}$.

\begin{figure}
    \centering
        \includegraphics[width=0.6\textwidth]{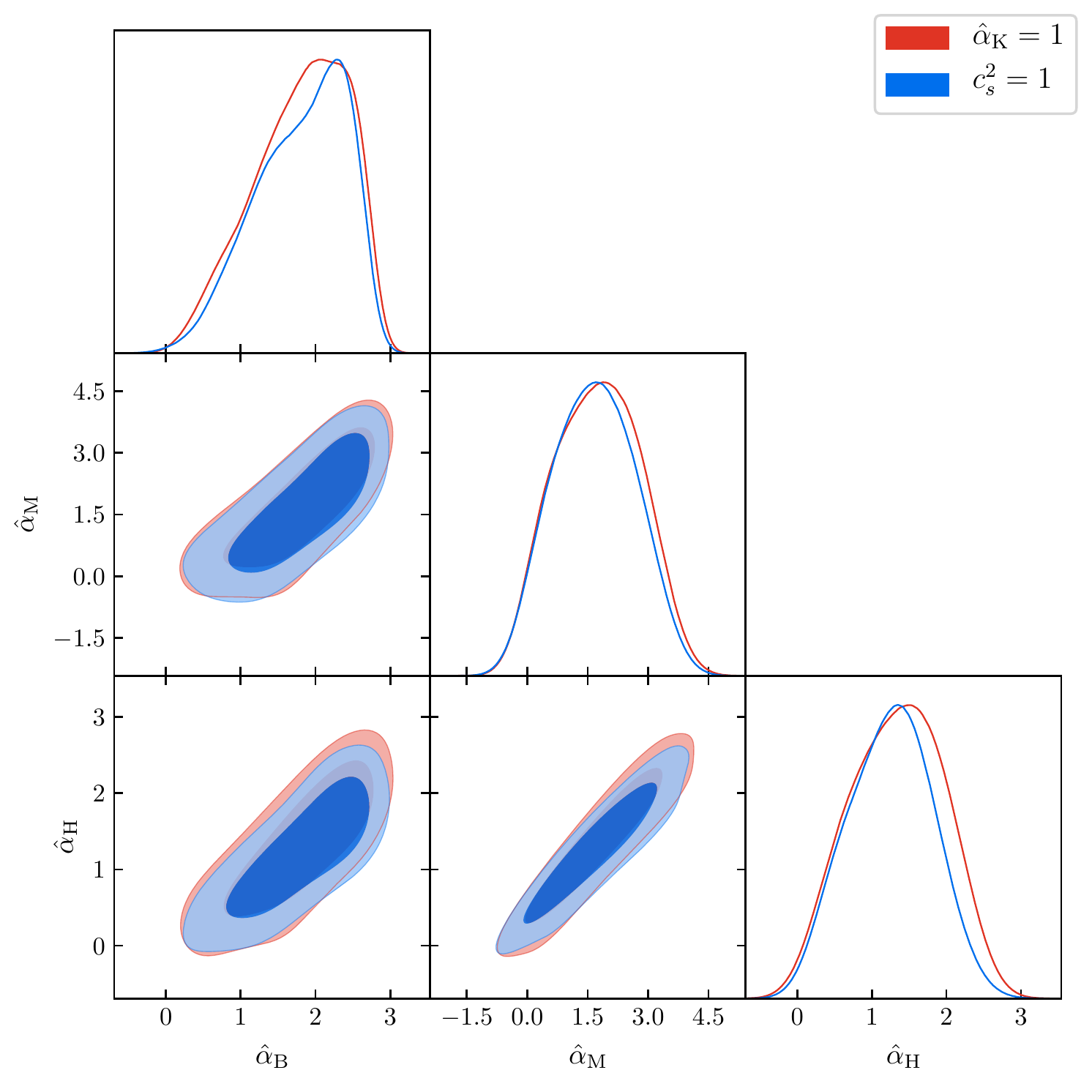}
     \caption{Constraints on $\hbra$, $\hrun$ and $\hbeh$ from the combination of CMB+BAO+RSD datasets. The background expansion history is again $\Lambda$CDM and the evolve the $\alpha_i$ is proportional to $\Omega_{\rm DE}$, with $\hkin =1$ (red), and $\cs=1$ and $\kin$ given by Eq. (\ref{equ:kin_fix}) (blue). We also vary the standard cosmological parameters.}
     \label{fig:ak4}
\end{figure}
On Figure~\ref{fig:ak4} we show the different constraints on the parameters for the two cases discussed in Section~\ref{sec:gw17}. 
The combination of datasets used for both contours here is CMB+BAO+ RSD and the model considered is $\Lambda$CDM+$\alpha_{BMH}$.
The red contour represents the case where we set $\hkin=1$ and let the speed of sound of the scalar degree of freedom vary with the other $\hat{\alpha}_i$.
The blue is for the case where the speed of sound of the scalar $\cs=1$ and $\kin$ is determined by the relation in Eq.~(\ref{equ:kin_fix}).
We can see that there is no significant difference in the contours for those two cases, which confirms that the choice for $\kin$ has little effect on current constraints, and that the $\cs=1$ branch of the results of \cite{Creminelli:2018xsv} may not have an effect on the estimated values of the rest of the $\alpha_i$.

\section{Discussion}\label{sec:discussion}

In this paper, we studied the phenomenology of the beyond Horndeski class of theories.
This class of theories includes the most popular DE and MG models that aim at explaining the accelerated expansion of the Universe and has been shown to have self-accelerating solutions.
The minimal set of functions that can describe the the evolution of linear perturbations in this model is $\{\kin,\bra,\run,\ten,\beh\}$.
To find the solutions of these equations a parametric form for these functions has to be assumed.
Here we chose the time evolution to be given by the fractional density of dark energy, so that $\alpha_i = \hat{\alpha}_i\,\Omega_{\rm DE}$.
Taking into account the implications from the observation of the binary neutron star merger (GW170817) and its electromagnetic counterpart (GRB170817A), we have set the tensor speed excess, $\hten = 0$.

Constraints on $\hkin$, $\hbra$, $\hrun$ and $\hten$ from cosmological data are $\mathcal{O}(1)$.
In this work our aim was to constrain the parameter $\hbeh$, which is peculiar of the theories beyond Horndeski, using a combination of CMB, BAO and RSD datasets.

Studying the phenomenology of this model and in particular the case where $\hkin=1$, $\hbra=\hrun=\hten=0$, we found that $\hbeh$ suppresses the matter power spectrum and the CMB lensing potential, and enhances the temperature power spectrum of the CMB at low $l$.
These findings are consistent with other results in the literature.
The quasi-static approximation was used to derive analytic solutions to evolution equations of the matter overdensity, $\delta_{\rm m}$ the growth rate parameter, $f$ and the weak lensing parameter, $\Sigma$, which agreed with the exact solutions obtained using \texttt{hi\_class}.
We discuss the commonly used notation for the quasi-static approximation and suggest how that may be amended to account for the extra terms that come from the presence of the beyond Horndeski parameter, $\beh$.

We perform parameter estimation using CMB temperature, polarization and lensing data from the Planck 2015 dataset and measurements of the BAO and RSD from a number of different large scale structure surveys. 
From the MG parameters we set $\hkin=1$, as varying this parameter has been shown to have no effect on the other parameters, as mentioned above we also set $\hten=0$, and varied $\hbra$, $\hrun$ and $\hbeh$.
We performed different runs where we varied $\hbeh$ and the six standard cosmological parameters, $h_0$, $\Omega_{\rm b}$, $\Omega_{\rm cdm}$,  $A_s$, $n_s$ and $\tau$, combined with either $\Omega_\nu$, $\hbra$ and $\hkin$, $w_0$ and $w_a$, or a combination of all.
We present the confidence contours for two of these runs and show that the beyond Horndeski parameter, $\hbeh$, is degenerate with the other MG parameters, $\hbra$ and $\hrun$, but not with either the $\Lambda$CDM parameters, $\Omega_\nu$, $w_0$ or $w_a$.
The constraints on $\hbeh$ that we get from these runs are $\mathcal{O}(1)$.
We also show that setting the speed of sound of the scalar $\cs=1$ does not have a substantial effect on the results.
Comparing the likelihood of the best fit models for the different runs we have done, we find that the MG models seem to fit the data slightly better than $\Lambda$CDM.
However, the difference in the likelihood is too small to comment on the validity of either model over the others and most probably is simply due to the fact that there are more free parameters in the MG models than there are in $\Lambda$CDM.

\section*{Acknowledgements}
We are very grateful to Joe Zuntz for helping us in the implementation of \texttt{hi\_class} in \texttt{CosmoSIS}. We thank Erminia Calabrese for useful comments and discussions. We are also grateful to Marco Crisostomi and Guillem Domenech for useful comments on this manuscript. DT, EB and PGF are supported by ERC H2020 693024 GravityLS project, the Beecroft Trust and STFC.

\newpage
\bibliography{beyond_horndeski}
\bibliographystyle{vancouver}
\newpage
\appendix
\section{Implementing the equations into \texttt{hi\_class}}\label{app:hi-class}
In this section we present the equations we implemented into \texttt{hi\_class}. In CLASS (and \texttt{hi\_class}) the units are chosen so that the Friedmann equations are written as,
\begin{align}
&H^2 = \rhomc + \rhosc\,,\\
&\frac{H^{\prime}}{a} = -\frac{3}{2} \left(H^2 + \pmc + \psc\right)\,,\\
\end{align}
where the primes denote a derivative with respect to conformal time, however the Hubble parameter $H$ is the physical one, $\rhomc$ and $\pmc$ are the density and pressure of matter in the universe (excluding dark energy), and $\rhosc$ and $\psc$ are those of the scalar field.
\begin{equation}
    \begin{aligned}
         \rhosc \equiv& - \frac{1}{3} K + \frac{2}{3} X \left(K_{X} - G_{3\phi}\right) -  \frac{2 H^3 \phi^{\prime} X}{3 a}\left[7G_{5X} + 4 X \left(G_{5X X} - 33F_5 - 12 X F_{5X}\right)\right]\\
         &+H^2\left[1 - \left(1-\bra\right)\Ms - 4X\left(G_{4X} - G_{5\phi}\right) - 4X^2\left(2G_{4XX} - G_{5\phi X} + 8 F_4 + 4X F_{4X}\right)\right]\,,
    \end{aligned}
\end{equation}

\begin{equation}
    \begin{aligned}
        \psc =& \frac{1}{3} K - \frac{2}{3} X \left(G_{3\phi} - 2G_{4\phi \phi}\right) + \frac{4 H \phi^{\prime}}{3 a}\left(G_{4\phi} - 2X G_{4\phi X} + X G_{5\phi \phi} - 4X^2 F_{4\phi}\right) -  \frac{\left(\phi^{\prime\prime} + a H \phi^{\prime}\right)}{3 \phi^{\prime} a}  H \Ms\bra \\
        & - \frac{4}{3} H^2 X^2 \left(G_{5\phi X} - 12 X F_{5\phi}\right) - \left(H^2 + \frac{2 H^{\prime}}{3 a}\right)\left(1-\Ms\right) + \frac{2 H^3 \phi^{\prime} X}{3 a}\left(G_{5X} - 12 X F_5\right)\,.
    \end{aligned}
\end{equation}
This form of the densities and pressures is analogous to the way they were defined in \cite{Zumalacarregui:2016pph}, so that we get the usual conservation equations,
\begin{align}
&\rhomc^{\prime} = - 3  H a \left(\rhomc + \pmc \right)\,,\\
&\rhosc^{\prime} = - 3  H a \left(\rhosc + \psc \right) \,.
\end{align}
In CLASS it is possible to solve the linear perturbation equations in both Newtonian and synchronous gauge.
However, the current version of \texttt{hi\_class} only has the synchronous gauge, so here we present the equations of Section~\ref{sec:perts} in this gauge as they are implemented into the code.\\
The line element to first order in the perturbations in synchronous gauge is \cite{Ma:1995ey},
\begin{equation}
ds^2=a^2\left[-d\tau^2+\left(\delta_{ij}+\tilde{h}_{ij}\right)dx^idx^j\right]\,,
\end{equation}
with
\[
\tilde{h}_{ij}(\tau,\Vec{k}\,)=\hat{k}_i\hat{k}_jh+6\left(\hat{k}_i\hat{k}_j-\frac{1}{3}\delta_{ij}\right)\eta+h_{ij}\,,
\]
in Fourier space, where $h$ and $\eta$ are the scalar perturbations and $h_{ij}$ is the tensor perturbation. \\
The scalar field perturbation $V_{\rm X}$ here is defined as 
\begin{equation}
V_{\rm X}\equiv a\frac{\delta\phi}{\phi^{\prime}}= a\,v_{\rm X}\,.
\end{equation}
In this gauge the Einstein equations take the form:

Einstein (0,0)
\begin{align*}
h^{\prime} =&\, \frac{4 k^2}{H a} \left(\frac{1 + \beh}{2 - \bra}\right) \eta + \frac{6 \rhomc \delta_{\rm mc} a}{H \Ms \left(2 - \bra\right)} - 2 H a \left(\frac{ 3 \bra + \kin} {2 - \bra}\right) V_X^{\prime}\\
& - 2 \left[3 a H^{\prime} + \left(\frac{\kin + 3 \bra}{2 - \bra}\right)a^2 H^2 + \frac{9 a^2}{\Ms}\left(\frac{\rhomc + \pmc} {2 - \bra}\right) + \left(\frac{\bra + 2 \beh} {2 - \bra}\right)k^2\right]V_X\,.
\end{align*}

Einstein (0,i)
\begin{equation}
\begin{aligned}
\eta^{\prime} = \frac{3 a^2 \theta_{\rm m}}{2 k^2 \Ms} + \frac{1}{2} \bra H a V_X^{\prime} + \left[ H^{\prime} a + \frac{1}{2} \bra H^2 a^2 + \frac{3 a^2}{2 \Ms}\left(\rhomc + \pmc\right)\right] V_X\,.
\end{aligned}
\end{equation}
where 
\[\theta_{\rm m}\equiv - \left(\rhomc + \pmc\right) \frac{k^2}{a} v_m\,.\]

Einstein (i,j) traceless
\begin{equation}
\begin{aligned}
 \xi^{\prime} = \left(1 + \ten\right) \eta - a H \left(2 +  \run\right) \xi + a H \left( \run -  \ten - \beh \right) V_X  - \beh V_X^{\prime}  - \frac{9 a^2 \sigma_{\rm m}}{2 k^2 \Ms}\,.
\end{aligned}
\end{equation}
To simplify this equation a new perturbation $\xi(\tau,\vec{k})$ has been introduced that relates $h$ and $\eta$ through
\[
\xi=\frac{h^{\prime}+6\eta^{\prime}}{2k^2}\,.
\]

Einstein (i,j) trace
\begin{equation}
\begin{aligned}
D h^{\prime\prime} =&\, 2 \left(\lambda_1 - \lambda_{9}\right) k^2 \eta - \frac{6 \bra \beh k^2 \eta^{\prime}}{H a} + 2 a H \lambda_3 h^{\prime} -  \frac{9 \kin \delta \pmc a^2}{\Ms}\\
&  + \left(3 a^2 H^2 \lambda_4 -2 \beh \kin k^2\right) V_X^{\prime} + 2 a H k^2 \left(\lambda_{10} + \lambda_5 + 3 \frac{H^2 a^2}{k^2} \lambda_6\right)  V_X\,,
\end{aligned}
\end{equation}

And the equation of the evolution of the scalar perturbations is given by
\begin{equation}
\begin{aligned}
&D\left(2 -  \bra\right) V_X^{\prime\prime} + \left[\frac{ \bra \beh k^2}{H a}\left(2-\bra\right) + 8 H a \lambda_7\right] V_X^{\prime} + \left[2 k^2 \lambda_{11} + 2 \csn k^2 - 8 H^2 a^2 \lambda_8\right] V_X \\
&= \frac{2 k^2}{H a} \left(\csn + \frac{3 (\rhomc+\pmc)}{H^2 M_*^2} \beh(1+\beh)\right) \eta + \frac{2 \beh k^2}{H^2 a^2} \left(2 - \bra\right) \eta^{\prime} + \frac{3 a }{2 H M_*^2}\left[ 2 \lambda_2 \delta\rhomc - 3 \bra \left(2-\bra\right) \delta \pmc\right] \,.
\end{aligned}
\end{equation}
where 
\begin{align*}
&D = \kin + \frac{3}{2}\bra^2\,,\\
&\lambda_1 = \kin(1+\ten) - 3\bra(\run-\ten)\,,\\
&\lambda_2 = -\frac{3(\rhomc + \pmc)}{H^2\Ms} - (2 - \bra)\frac{H^{\prime}}{aH^2} + \frac{\bra^{\prime}}{aH}\,,\\
&\lambda_3 = -\frac{1}{2}(2+\run)D-\frac{3}{4}\bra\lambda_2\,,\\
&\lambda_4 = \kin\lambda_2 - \frac{2\kin\bra^{\prime} - \bra\kin^{\prime}}{aH}\,,\\
&\lambda_5 = \frac{3}{2}\bra^2(1+\ten)+(D+3\bra)(\run - \ten) + \frac{3}{2}\bra\lambda_2\,,\\
&\lambda_6 = \left(1-\frac{3\bra H^{\prime}}{\kin a H^2}\right)\frac{\kin\lambda_2}{2}-\frac{D H^{\prime}}{aH^2}\left(2+\run + \frac{H^{\prime\prime}}{a H H^{\prime}}\right) - \frac{2\kin\bra^{\prime}-\bra\kin^{\prime}}{2aH} - \frac{3\kin\pmc^{\prime}}{2aH^3\Ms}\,,\\
&\lambda_7 = \frac{D}{8}\left(2-\bra\right)\left(4+\run+\frac{2H^{\prime}}{aH^2}+\frac{D^{\prime}}{aHD}\right) + \frac{D}{8}\lambda_2\,,\\
&\lambda_8 = -\frac{\lambda_2}{8}\left(D-3\lambda_2 + \frac{3\bra^{\prime}}{aH}\right) + \frac{1}{8}(2-\bra)\left[\left(3\lambda_2-D\right)\frac{H^{\prime}}{aH^2}-\frac{9\bra\pmc^{\prime}}{2aH^3\Ms}\right]\\
&\qquad- \frac{D}{8}\left(2-\bra\right)\left[4+\run+\frac{2H^{\prime}}{aH^2}+\frac{D^{\prime}}{aHD}\right]\,,\\
&\lambda_{9} =  3\left[\bra \beh\left( 1 + \run\right) +  \frac{ \bra \beh^{\prime}}{a H}\right]\,,\\
&\lambda_{10} = \lambda_9 + 3 \bra \beh \frac{H^{\prime}}{aH^2} - \beh \kin\,,\\
&\lambda_{11} =\beh\left[\frac{1}{2}(2-\bra) \left(\bra + \frac{2H^{\prime}}{a H^2}\right) + \frac{3 (\rhomc+\pmc)}{H^2 M_*^2} (1+\beh)\right]\,,\\
&c_{\rm sN}^2 = \frac{1}{2}(2-\bra)\left[\bra + 2\beh +2\run(1+\beh) - (2-\bra)\ten + \frac{2\beh^{\prime}}{a H} - 2\frac{(1+\beh) H^{\prime}}{a H^2}\right]\\
&\qquad+ \frac{(1+\beh)\bra^{\prime}}{a H} - \frac{3 (\rhomc+\pmc)}{H^2 M_*^2} (1+\beh)^2\,,
\end{align*}
where $c_{\rm sN}^2$ is the numerator of the speed of sound of the scalar,
\[
c_s^2=\frac{c_{\rm sN}^2}{D}\,.
\]
To include the effects of beyond Horndeski theories, we added the terms involving $\beh$, $\lambda_9$, $\lambda_{10}$ and $\lambda_{11}$ to the speed of sound of the scalar and perturbation equations and the $F_i$ and their derivatives to the definitions of the scalar density and pressure and the $\alpha_i$ in \texttt{hi\_class}.\\
For now we have only included two parametrisations for this model in \texttt{hi\_class}, one where the evolution of $\alpha_i$ is given by the evolution of the DE density parameter $\Omega_{\rm DE}$ and and one where they are proportional to the scale factor,
\begin{align}
    &\alpha_i=\left(1-\Omega_{\rm m}\right)\hat{\alpha}_i\,,\\
    &\alpha_i= a\,\hat{\alpha}_i\,.
\end{align}
For both of these you need to set or vary the initial set $\{\hkin,\hbra,\hrun,\hten,\hbeh\}$.
For the first one we have also added the option to have the value of $\kin$ computed using Eq.~(\ref{equ:kin_fix}).
In this case the initial set of parameters you need to specify is $\{\hbra,\hrun,\hten,\hbeh\}$.
\end{document}